\documentclass[12pt]{amsart}
\usepackage{amsmath,amsthm,amsfonts,amssymb,eucal, hyperref}

\newcommand{\Om}{\Omega}

\newcommand{\oq}{\ {\raise 7pt\hbox{${\scriptstyle\circ}$}}
\kern -7pt{%\lower 2pt
\hbox{$Q$}}}

\newcommand{\R}{ \mathbb R}
\newcommand{\N}{ \mathbb N}

\newcommand {\BR}{\mathbf R}

\newcommand {\be}{\mathbf e}

\newcommand{\lu}{\langle}
\newcommand{\ru}{\rangle}

\newcommand{\CS}{\mathcal S}

\newcommand{\meas}{\mathrm {meas}}

\newcommand{\1}
{{\,\vrule depth3pt height9pt}{\vrule depth3pt height9pt}
{\vrule depth3pt height9pt}{\vrule depth3pt height9pt}\,}

\DeclareMathOperator{\op}{{Op}}

\DeclareMathOperator {\ad}{{ad}}
\DeclareMathOperator{\supp}{{supp}}

\newtheorem{thm}{Theorem}[section]
\newtheorem{cor}[thm]{Corollary}
\newtheorem{cla}[thm]{Claim}
\newtheorem{lem}[thm]{Lemma}

\theoremstyle{definition}
\newtheorem{rem}[thm]{Remark}

\numberwithin{equation}{section}

\newcommand{\bee}{\begin{equation}}
\newcommand{\ene}{\end{equation}}
\newcommand{\bees}{\begin{equation*}}
\newcommand{\enes}{\end{equation*}}
\newcommand{\bes}{\begin{split}}
\newcommand{\ens}{\end{split}}

\newcommand{\bet}{\begin{thm}}
\newcommand{\ent}{\end{thm}}
\newcommand{\bel}{\begin{lem}}
\newcommand{\enl}{\end{lem}}
\newcommand{\bec}{\begin{cor}}
\newcommand{\enc}{\end{cor}}
\newcommand{\becl}{\begin{cla}}
\newcommand{\encl}{\end{cla}}
\newcommand{\bep}{\begin{proof}}
\newcommand{\enp}{\end{proof}}
\newcommand{\ber}{\begin{rem}}
\newcommand{\enr}{\end{rem}}
\newcommand{\ep}{\varepsilon}
\newcommand{\la}{\lambda}

\newcommand{\de}{\delta}
\newcommand{\al}{\alpha}
\newcommand{\Z}{\mathbb Z}

\newcommand{\De}{\Delta}

\makeatletter
\def\square{\RIfM@\bgroup\else$\bgroup\aftergroup$\fi
  \vcenter{\hrule\hbox{\vrule\@height.6em\kern.6em\vrule}\hrule}\egroup}
\makeatother

\usepackage[letterpaper,hmargin=1in,vmargin=1in]{geometry}
\usepackage{graphicx,mathtools,bm,tabu}
\begin{document}

\hoffset -4pc

\title[Perturbation Theory]
{Perturbation theory for almost-periodic potentials I. One-dimensional case.}
\author[L. Parnovski \& R. Shterenberg]
{Leonid Parnovski \& Roman Shterenberg}
\address{Department of Mathematics\\ University College London\\
Gower Street\\ London\\ WC1E 6BT\\ UK}
\email{Leonid@math.ucl.ac.uk}
\address{Department of Mathematics\\ University of Alabama at Birmingham\\ 1300 University Blvd.\\
Birmingham AL 35294\\ USA}
\email{shterenb@math.uab.edu}

%\keywords{Periodic operators, almost-periodic pseudodifferential operators, integrated density of states}
%\subjclass[2000]{Primary 35P20, 47G30, 47A55; Secondary 81Q10}
%\thanks{This work is supported by the Royal Society}
%\copyrightinfo{2002}{Alexander V. Sobolev}

%\begin{abstract}
%\input polyabstract.tex
%\end{abstract}
%\begin{abstract}
%We consider a periodic pseudodifferential operator $H=(-\Delta)^l+B$ ($l>0$) in $\R^d$ which satisfies the
%following conditions: (i) the symbol of $H$ is smooth in $x$, and (ii) the perturbation
%$B$ has order smaller than $2l$. Under these assumptions, we prove that the spectrum
%of $H$ contains a half-line.

%spectral gaps}
%\end{abstract}
\date{\today}

\begin{abstract}
We consider the family of operators  $H^{(\ep)}:=-\frac{d^2}{dx^2}+\ep V$ in $\R$ with almost-periodic potential $V$.  
We study the behaviour of the integrated density of states (IDS) $N(H^{(\ep)};\la)$ when $\ep\to 0$ and $\la$ is a fixed energy. When $V$ is quasi-periodic (i.e. is a finite sum of complex exponentials), we prove that for each $\lambda$ the IDS has a complete asymptotic expansion in powers of $\epsilon$; these powers are either integer, or in some special cases half-integer. These results are new even for periodic $V$. We also prove that when the potential is neither periodic nor quasi-periodic, there is an exceptional set $\CS$ of energies (which we call {\it the super-resonance set}) such that for any $\sqrt\la\not\in\CS$ there is a complete power asymptotic expansion of IDS, and when $\sqrt\la\in\CS$, then even two-terms power asymptotic expansion does not exist. We also show that the super-resonant set $\CS$ is uncountable, but has measure zero. Finally, we prove that the length of any spectral gap of $H^{(\ep)}$ has a complete asymptotic expansion in natural powers of $\ep$ when $\ep\to 0$.
\end{abstract}

\

\maketitle
\vskip 0.5cm

\section{Introduction}

We consider the operator 
\bee
H=H^{(\ep)}:=-\De+\ep V,
\ene
where $\ep>0$ is a small parameter and $V$ is a real-valued almost-periodic potential. 
We are interested in various quantitative and qualitative spectral properties of $H$ as $\ep\to 0$, and this paper is the first one in a series of articles devoted to the study of these properties of $H$ %in different settings. 
under various assumptions. In this paper we assume that the dimension $d=1$, so that 
\bee\label{1.2}
H=H^{(\ep)}:=-\frac{d^2}{dx^2}+\ep V.
\ene
The quantities we will be interested in are: the length of the spectral gaps, and  the integrated density of states (IDS) $N(\la;H^{(\ep)})$ when the spectral variable $\la$ is fixed; we are going to study the behaviour of these quantities as $\ep\to 0$.

The first problem we study is as follows. Let $\la\in\R$  be a fixed number and consider the behaviour of the IDS of $H^{(\ep)}$ at $\la$ when $\ep\to 0$. Questions of this nature (how the value of IDS at a fixed energy depends on the value of a small coupling constant) have arisen in our study 
of perturbations of Landau Hamiltonians by almost-periodic potentials. Despite the slightly esoteric feel of this type of questions, we believe they are more natural than it may seem at the first sight, especially given that the answers are quite surprising. Let us briefly describe the effects happening in dimension one; we are going to devote the second paper in this series to discuss  the  multidimensional case, where the results are even more unexpected. Suppose first that $V$ is quasi-periodic (i.e. $V$ is a finite linear combination of trigonometric functions). Then, whenever $\la$ is not a square of a frequency, there is a complete asymptotic expansion of $N(\la;H^{(\ep)})$ in integer powers of $\ep$. Suppose, $\la=\theta^2\ne 0$, where $\theta$ is a frequency. Then the type of the expansion will depend on the relationship between $\tau$ (the constant Fourier coefficient of $V$) and $\nu$ (the Fourier coefficient  at $e^{i2\theta x}$). First we notice that, as we will show in this paper,  there is a spectral gap of  $H^{(\ep)}$ around $\theta^2$ of length $\sim 2\nu\ep$. Therefore, if $|\tau|<|\nu|$, then the point $\la+\tau\ep$ stays inside this gap and, as a result, the IDS does not depend on $\ep$ when $\ep$ is small. If, on the other hand, $|\tau|>|\nu|$, then the shift by $\tau\ep$ pushes our point $\la$ well outside the spectral gap, and we obtain the standard asymptotic expansion in integer powers of $\ep$. The most interesting case is $|\tau|=|\nu|$, when the point $\la+\tau\ep$ is approximately at the edge of the spectral gap. In this case generically the answer will depend on the sign of $\tau$. For one value of this sign the point $\la+\tau\ep$ is still located in the gap and so the IDS is constant. However, for the opposite value of the sign of $\tau$  
the point $\la+\tau\ep$ will be pushed just outside the gap and, as a result, the IDS will have a complete expansion in half-integer powers of $\ep$ (where we define half-integers as $(\Z/2)\setminus\Z$). Similar situation happens when we look at the point $\la=0$: we have expansion in half-integers whenever $\tau<0$; otherwise, the expansion is in integers. 
The bottom line is, if $V$ is quasi-periodic, then for all $\la$ we have a complete asymptotic expansion of $N(\la;H^{(\ep)})$ as $\ep\to 0$, which contains either integer, or half-integer powers of $\ep$. 

An interesting phenomenon occurs when we look at this problem in the  `proper' almost-periodic setting, for example, when all the Fourier coefficients  are non-zero. 
In this case there is a substantial set $\CS$ such that for $\la^{1/2}\in\CS$ there is no asymptotic expansion of $N(\la;H^{(\ep)})$ at all; in fact, there are uncountably many values of $\la$ for which the remainder $N(\la;H^{(\ep)})-N(\la;H^{(0)})$ is not even  asymptotically equivalent to any power of $\ep$. This set (which we call the {\it super-resonance set}) is therefore uncountable, but has measure zero; the interesting feature of this set is that it is present no matter how quickly the Fourier coefficients of $V$ go to zero -- whether $V$ is smooth, or analytic, the super-resonant set without the asymptotic expansion of IDS is always uncountable (but perhaps its dimension may depend on the smoothness of $V$).  
%This result therefore is a counterexample to a wide-spread `folk philosophy' in the subject `Whichever results one can prove in the quasi-periodic case, they  

The second problem we consider is as follows.
It has been noticed by Arnold, \cite{A} that if $H$ is a Hill operator \eqref{1.2} with $V$ being a finite trigonometric {\bf periodic} polynomial 
\bee\label{Ar}
V=\sum_{j=-n}^na_je^{ i2 jx},
\ene
%with $a_{-j}=\bar{a_j}$
then the size of the spectral gap around the point $m^2$, $m\in\N$ is at most $C_{m,n}\ep^{-[-m/n]}$, where $[s]$ is the largest integer not bigger than $s$. 
If the sum in \eqref{Ar} is infinite, then the size of any gap is (generically) proportional to $\ep$. It turns out that in the periodic case it is not very difficult to prove more: that the size of each spectral gap enjoys a complete asymptotic expansion in natural powers of $\ep$, see e.g. \cite{Bor}. Our second theorem is the extension of these results  to the case of almost-periodic potentials: we prove that the length of each spectral gap has a complete asymptotic expansion in natural powers of $\ep$. We also prove similar expansions for the upper and lower ends of each spectral gap. The leading power in each expansion will depend on whether the potential is quasi-periodic or almost-periodic. In the quasi-periodic case the leading power of the length of the gap opened around the square of each frequency $\theta$ will increase together with the order of $\theta$   
 (see the next section for the precise definitions and formulation of the results), whereas in the almost-periodic setting when no Fourier coefficients vanish, all expansions begin with the first power of $\ep$. 
These expansions are formally uniform, but effectively they are not, because the higher the order of a frequency $\theta$ is, the smaller $\ep$ we need to choose to `see' the expansion of the length of the  gap generated by $\theta$ (i.e. if we choose $\ep$ not very small, then the remainder in the expansion will be larger than the asymptotic terms). 

Somewhat similar problems were considered in \cite{DaGo} and, in the discrete setting, in \cite{Le} (see also \cite{LiSh} and references there). However, there is a significant difference between these papers and our results. In these papers the authors have fixed $\ep$ and studied the behaviour of the gap length as a function of the `natural label' of the gap (corresponding, roughly, to what we call the order of a frequency, see below for details). So, they were able to obtain information about all gaps simultaneously, but this information was either bounds (upper and lower), or one asymptotic term, whereas we obtain more detailed information (complete asymptotic expansion) about smaller number of gaps.

The method we use for obtaining our results is a version of the gauge transform method used in \cite{ParSob} and \cite{PaSh1}. The only difference is that in  \cite{ParSob} and \cite{PaSh1} we had fixed coupling constant and assumed that the energy $\lambda$ was large (so that the small parameter was $\la^{-1}$), whereas in the present paper the small parameter is the coupling constant $\ep$. This difference is not essential, so the construction of the gauge transform can be performed almost word-to-word as it is done in \cite{ParSob} and \cite{PaSh1}. This method allows us to find two operators, $H_1$ and $H_2$ so that $H_1$ is unitarily equivalent to $H$, $H_2$ is close to $H_1$ in norm, and $H_2$ is almost diagonal (in the sense that most of the off-diagonal matrix coefficients of $H_2$ vanish). 
For the sake of completeness, we have written the details of the gauge transform construction relevant to our setting in the Appendix; in the main body of the paper we will give a brief description of the method and use the relevant properties of $H_1$ and $H_2$ without proof. 

The structure of the rest of the paper is as follows: in the next section we will give all the necessary definitions and formulate the main results. In Section 3 we will discuss the quasi-periodic operators, and in Section 4 the almost-periodic operators. Finally, in the Appendix we will describe the method of the gauge transform. 

\subsection*{Acknowledgments}
We are grateful to Alexander Sobolev for reading the preliminary version of this manuscript and making useful suggestions. 
We are grateful to the referee for useful comments. 
The research of the first author was partially supported by the EPSRC grant EP/J016829/1.

\section{Notation and main results}

%\subsection{quasi-periodic results}
We will consider two types of the potential $V$. The first type  is quasi-periodic potentials:
\bee\label{quasi}
V=\sum_{\theta\in\Theta}\hat V_{\theta}\be_{2\theta}.
\ene
Here, 
\bee
\be_{\theta}( x):=e^{i\theta x},
\ene
$\hat V_{\theta}$ are complex numbers (called the Fourier coefficients of $V$; since $V$ is real, we have $\hat V_{-\theta}=\overline{\hat V_{\theta}}$), and $\Theta=\Theta(V)\subset\R^d$ is a finite set, called the set of  frequencies (or rather half-frequencies; the factor $2$ is used purely for convenience) of $V$. %(and taking the exponent to be $2\theta x$, rather than $\theta x$ is done purely for convenience). 
We assume without loss of generality that $\Theta$ is symmetric about the origin and contains it. Denote by $l$ the number of independent elements in $\Theta$ (so that $|\Theta|=2l+1$).  
For each natural $L$ we denote $\Theta_L:=\Theta+\Theta+\dots+\Theta$ (the algebraic sum of $L$ copies of $\Theta$) and put $\Theta_{\infty}:=\cup_L\Theta_L$. When 
$\theta\in\Theta_{\infty}$, we denote by $Z(\theta)$ the smallest number $L$ for which $\theta\in\Theta_L$ and call this number the {\it order} of the frequency $\theta$. We  put 
\bee
T_L:=\Theta_L\setminus\Theta_{L-1}. 
\ene
A simple combinatorial estimate shows that 
\bee\label{number}
\#\Theta_L\le %{L\choose  3l}\ll_l 
(3L)^{3l}. 
\ene
We also put $\tau:=\hat V_0$, $\Theta':=\Theta\setminus \{0\}$ and $V':=V-\tau$, so that
\bee
V'=\sum_{\theta\in\Theta'}\hat V_{\theta}\be_{2\theta}.
\ene

%\subsection{almost-periodic results}
%We call such potentials smooth almost-periodic, although this class is smaller than the usual set of almost-periodic %functions, since we still assume that $\Theta$ is finite. Obviously, $V$ is periodic iff $\Theta_{\infty}$ is discrete. 

The second type of potentials we are going to consider are smooth almost-periodic, by which we mean that  $\Theta$ is still a finite set, but we have 
\bee
V=\sum_{\theta\in\Theta_{\infty}}\hat V_{\theta}\be_{2\theta}
\ene
with 
\bee
|\hat V_{\theta}|\ll m^{-P}
\ene
for $\theta\in T_m$ and arbitrary positive $P$. %recall that
%\bee
%T_m:=\Theta_m\setminus\Theta_{m-1}. 
%\ene
 
We also assume that $\Theta$ satisfies the diophantine condition, i.e. for $\theta\in \Theta_m$ we have $|\theta|\gg m^{-P_0}$, where $P_0>0$ is fixed. 

In either of these two cases (quasi- or almost-periodic potentials) we also assume (as we can do without loss of generality) that 
\bee\label{L2V}
||V||_2:=(\sum_{\theta\in\Theta_{\infty}}|\hat V_{\theta}|^2)^{1/2}<1/100.
\ene

Our first main result concerns the spectral gaps.

\bet \label{gapst1int} Suppose, $V$ is either quasi-periodic, or infinitely smooth almost-periodic and satisfies all the above assumptions.  Suppose, $\theta\in\Theta_{\infty}'$. Then for sufficiently small $\ep$ operator $H$ has a (possibly trivial) spectral gap around $|\theta|^2$, the length of which, as well as its upper and lower ends, have  complete asymptotic expansions in natural powers of $\ep$. If $\hat V_\theta\not=0$ then the asymptotic expansion for the upper (lower) end of the gap starts with $|\theta|^2\pm |\hat V_\theta|\varepsilon +O(\varepsilon^2)$.
%If $\hat h_2(-\theta_0;2\theta_0)\equiv 0$ for each $N$, then the size of the spectral gap around $|\theta|^2$ is $O(\varepsilon^{\infty})$. If $\hat h_2(-\theta_0;2\theta_0)\not\equiv 0$ for some $N$, 
\ent
\ber
If $\hat V_{\theta}=0$, we cannot guarantee that an expansion for the gap-length is always non-trivial, i.e. it could happen, in principle, that the length of the gap is $O(\ep^{+\infty})$. 
\enr

The next result involves two quantities, $s_2(0)$ and $g_2(0)$ which will be defined in the next section (in formula \eqref{notation}). Throughout the paper we use the convention that each time we use letters $a_j$ (or $a_j(\la)$) for coefficients in asymptotic expansions, the exact values of these coefficients  could be different. The same refers to the use of $C$ which can mean a different positive constant each time we use it. 

\bet\label{IDS}
Suppose, $V$ is quasi-periodic. Then for sufficiently small $\ep>0$ (i.e. $\ep<\ep_0$, where $\ep_0>0$ depends on $V$ and $\la$) the following holds:  

(i) For $\la<0$ we have $N(\la;H)=0$. 

(ii) For $\la>0$ and $\sqrt\la\not\in\Theta_{\infty}$ we have 
\bee\label{nonresas1bisi}
N(\la;H)\sim \pi^{-1}\sqrt\la+\sum_{p=1}^{\infty} a_{p}(\la) \ep^p.
\ene 

(iii) For $\la=|\theta_0|^2$ with $\theta_0\in\Theta'$ (i.e. $\nu:=\hat V_{\theta_0}\ne 0$) there are the following options:

(a) If either $|\tau|<|\nu|$, or $|\tau|=|\nu|$ and $[s_2(0)\tau-\Re (\nu\bar g_2(0))]<0$, then 
\bee
N(\la;H)=\pi^{-1}|\theta_0|; 
\ene
 
(b) If $|\tau|>|\nu|$, then 
\bee\label{eq:N2bisi}
N(\la;H)\sim \pi^{-1}|\theta_0|+\sum_{p=1}^\infty a_p(\la) \ep^p,\ \ a_1(\la)<0;
\ene 
 
(c) If $|\tau|=|\nu|$ and $[s_2(0)\tau-\Re (\nu\bar g_2(0))]>0$, then
\bee\label{eq:N3/2bisi}
N(\la;H)\sim \pi^{-1}|\theta_0|+\ep^{3/2}\sum_{p=0}^{\infty} a_p(\la) \ep^p,\ \ a_0(\la)<0;
\ene

(d) If $|\tau|=|\nu|$ and $[s_2(0)\tau-\Re (\nu\bar g_2(0))]=0$, then
\bee\label{eq:N3/2anew1i}
N(\la;H)\sim \pi^{-1}|\theta_0|+\ep^{k/2}\sum_{p=0}^{\infty} a_p(\la) \ep^p,\ \ a_0(\la)<0,
\ene 
with some natural $k\geq4$ including, possibly, $k=\infty$ (the latter means that $N(\la;H)= \pi^{-1}|\theta_0|+o(\ep^{+\infty})$). 
 
(iv) For $\la=|\theta_0|^2$ with $\theta\in\Theta_{\infty}\setminus\Theta$ (i.e. $\nu:=\hat V_{\theta_0}= 0$) there are the following options:
 
(a) If  $\tau=0$, and $s_2^2(0)<|g_2(0)|^2$, then 
\bee
N(\la;H)=\pi^{-1}|\theta_0|; 
\ene
 
(b) If $|\tau|>0$, then 
\bee\label{eq:N2bissi}
N(\la;H)\sim \pi^{-1}|\theta_0|+\sum_{p=1}^\infty a_p(\la) \ep^p,\ \ a_1(\la)<0.
\ene  

(c) If $\tau=0$ and $s_2^2(0)>|g_2(0)|^2$, then 
\bee\label{eq:N2bissi1}
N(\la;H)\sim \pi^{-1}|\theta_0|+\sum_{p=2}^\infty a_p(\la) \ep^p,\ \ a_2(\la)<0;
\ene  

(d) If $\tau=0$ and $s_2^2(0)=|g_2(0)|^2$, then
\bee\label{eq:N3/2anew2i}
N(\la;H)\sim \pi^{-1}|\theta_0|+\ep^{k/2}\sum_{p=0}^{\infty} a_p(\la) \ep^p,\ \ a_0(\la)<0,
\ene 
with some natural $k\geq5$ including, possibly, $k=\infty$.
 
(v) Suppose, $\la=0$. Then there are the following options:

(a) If $\tau>0$, then $N(0;H)=0$;

(b) If $\tau<0$, then 
\bee\label{lambda=0bissi}
N(0;H)\sim \ep^{1/2}(\pi^{-1}|\tau|^{1/2}+\sum_{p=1}^{\infty} a_p \ep^p);
\ene

(c) If $\tau=0$, then
\bee\label{lambda=01bisi}
N(0;H)\sim \ep\sum_{p=0}^{\infty} a_p \ep^p,\ \ a_0>0.
\ene
 
\ent

Finally, the following result holds for almost-periodic potentials. 

\bet
Suppose, $V$ is infinitely smooth %or analytic, 
almost-periodic, but not periodic, 
 and $\hat V_\theta\not=0$ for any $\theta\in\Theta'_\infty$. Then 
there exists a set $\CS$ (which we call a super-resonance set) such that a complete power asymptotic expansion of $N(\la;H)$ exists if and only if $\la\not\in\CS$.
The set $\CS$ is uncountable and has measure zero. 
\ent
\ber
As we will see in the proof, there are uncountably many values of $\la$ for which the difference 
$N(\la;H^{(\ep)})-N(\la;H^{(0)})$ properly oscillates between $C_1\ep^j$ and $C_2\ep^j$, where $C_1\ne C_2$ and $j$ equals $1$ or $2$.  
\enr

We will think of a point $\xi\in\R$ as the exponential function $\be_{\xi}( x):=e^{i\xi x}$ lying in the Besikovich space $B_2(\R)$ (the collection of all formal countable linear combinations of $\{\be_{\xi}\}$ with square-summable coefficients). Then for arbitrary pseudo-differential operator $W$ with symbol (in a left quantisation) $w=w(\xi,x)$ being quasi-periodic in $ x$,  
\bee
w(\xi,x)=\sum_{\theta\in\Theta}\be_{2\theta}( x)\hat w(\xi,\theta),
\ene
we have 
\bee
W\be_{\xi}=\be_{\xi}w(\xi,x).
\ene
Thus, we can think of the Fourier coefficients $\hat w(\theta,\xi)$ of the symbol as the matrix element of $W$ joining $\xi$ and $\xi+2\theta$:
\bee
\hat w(\xi,\theta)=\lu W\be_{\xi},\be_{\xi+2\theta}\ru_{B_2(\R)}.
\ene
In our paper \cite{PaSh1} it is explained that instead of working with operators acting in 
$L_2(\R)$, we can consider operators with the same symbol acting in $B_2(\R)$ and work with them. 
%we can work with operators acting in $L_2(\R)$ or $B_2(\R)$, since 
This will not change the spectral properties  we are studying in our paper (for example, the spectrum as a set is the same whether our operator acts in $L_2(\R)$ or $B_2(\R)$).

\section{Quasi-periodic potential}

In this section we assume that the potential $V$ is quasi-periodic, i.e. that \eqref{quasi} holds. 

\subsection{Gauge transform: general description} 

First of all, we give a brief outline of the construction of the gauge transform of our operator. The details of this construction are similar to those in \cite{PaSh1}; for the sake of completeness, we present them in the Appendix. Let us fix a natural number $N$. All the constructions are going to depend on the choice of $N$, but we will often omit writing $N$ as the variable. %We also assume, as we can without loss of generality, that $||V||_2=1$.  
%The construction of 
Applying the gauge transform leads to a pair of operators, $H_1=H_1^{(\ep)}$ and $H_2=H_2^{(\ep)}$ so that $H_1=UHU^{-1}$ is unitarily equivalent to $H$;  
$H_1$ and $H_2$ are close in norm, more precisely, 
\bee\label{eq:kand}
||H_1-H_2||\ll\ep^N; 
\ene 
and $H_2$ is almost diagonal in the sense that it can be decomposed into a direct integral with all fibres being finite dimensional (moreover, as we will see, the dimension of all fibres will be $1$ or $2$). Also, the frequencies of $H_2$ are inside the set $\Theta_{3N}$. Here, the coefficient $3$ technically appears in the gauge transform approach (see Appendix). It reflects the fact that one has to make slightly more than $N$ steps to achieve the error of order $\varepsilon^N$. Once we have constructed these operators, it turns out that we can study spectral characteristics of $H$ by means of studying the corresponding spectral characteristics of $H_2$. Indeed, the spectra of $H$ and $H_1$ are the same, and so are the lengths of the spectral gaps. Also, the lengths of the spectral gaps of $H_1$ and $H_2$ differ by at most $\ep^N$. 

Concerning the IDS, 
it was proved in \cite{PaSh1} that  
\bee\label{N}
N(\la;H_1)=N(\la;H) 
\ene
and 
\bee\label{eq:N1}
|N(\la;H_2)-N(\la;H_1)|\ll \ep^{(N+1)/2}. 
\ene

More precisely, we have shown in \cite{PaSh1} that the immediate consequence of \eqref{eq:kand} is 
\bee
N(\la-\ep^N;H_2) \le N(\la;H_1)\le N(\la+\ep^N;H_2), 
\ene  
and now \eqref{eq:N1} would follow once we establish expansion \eqref{eq:main_lem3} for  $N(\la; H_2^{(\ep)})$.

We also define 
\bee
H':=H_0+V'=H-\ep\tau, \ \ H_j':=H_j-\ep\tau, \ \ j=1,2,
\ene
and notice the obvious property
\bee
N(\la;H_j)=N(\la-\mu;H_j-\mu),
\ene
so if we put 
\bee
\mu:=\ep\tau,
\ene
we have
\bee
N(\la;H_j)=N(\la-\mu;H'_j).
\ene
This trivial consideration is important for understanding of some of the effects described later.

%Assume for simplicity at the moment that $V$ is quasi-periodic (i.e. $\theta$ is a finite set). 
Now we choose a small positive number $\de=\de(N)$, to be specified later %\footnote{describe the meaning of $\de$ here} 
and for each non-zero frequency $\theta\in\Theta'(H_2)$  
we put
\bee
R(\theta)=R(\theta;\delta):=\{\xi\in\R,\ |(\xi +2\theta)^2-\xi^2|<\delta\}
=(-\theta-\frac{\de}{4|\theta|},-\theta+\frac{\de}{4|\theta|}). 
%\ \ \ \delta=\delta(N).
\ene
Next, let $\psi=\psi(\xi)$ be a standard smooth non-negative cut-off function satisfying 
$\supp \psi \subset [-1/2,1/2]$ and $\psi(\xi)=1$ for $\xi\in[-1/4,1/4]$, and let   
$\varphi:=1-\psi$. 
We put 
\bee\label{phi}
\varphi_{\theta}(\xi):=\varphi((\xi+\theta)4|\theta|\de^{-1}).
%\begin{cases}
%0, & if \ \ \xi\in R(\theta) \\
%1, & \ otherwise
%\end{cases}
\ene
Note that 
\bee
R(-\theta)=-R(\theta), \quad \varphi_{-\theta}(-\xi)=\varphi_{\theta}(\xi).
\ene
We also put
$$
\tilde{\chi}_{\theta}(\xi):=\varphi_{\theta}(\xi)(|\xi+2\theta|^2-|\xi|^2)^{-1}=\frac{\varphi_{\theta}(\xi)}
{4(\xi+\theta)\theta}
$$
when $\theta\not= 0$, and $\tilde{\chi}_0(\xi)=0$. 

The region  $R(\theta)$ is called the resonance zone corresponding to $\theta$. Since (for fixed $N$) the number of resonance zones is finite and the length of them goes to zero, it implies that for sufficiently small $\delta$ these zones do not intersect.
%; this is the only condition we impose on the value of $\de$.\footnote{is it?} 
We also denote by
\bee
\BR(\delta):=\cup_{\theta\in\Theta'(H_2)}R(\theta;\delta)
\ene
the `overall' resonant set corresponding to $\ep$; we obviously have
\bee\label{BR}
\BR(\delta_1)\subset\BR(\delta_2) 
\ene
for $\delta_2>\delta_1$.
%for small $\ep_n$ it consists of two connected components 
%$R(\theta)^{\pm}$, where the sign in the superscript corresponds to the sign of the inner product $\lu\xi,\theta\ru$; we have $R(\theta)^{+}=R(-\theta)^{-}$.

In what follows we always assume that $\delta(N)$ is sufficiently small so that different resonance zones $R(\theta;\de)$ do not intersect for all $\theta\in\Theta'_{9N}$; we also take $\ep$ so small that $\ep\leq\delta^2$.

\ber\label{polynomialinN}
It is not difficult to see that in case when $\Theta'$ satisfies Diophantine condition on frequencies, the parameter $\delta(N)$ can be chosen to be $c^N$ with some constant $c=c(\Theta)$ with all constructions and statements of Section 3 being valid. %While this remark is not important for us here, such power dependence on $N$ may become essential for more subtle questions like description of the type of the spectrum of the operator. We plan to discuss it in our further publications.
\enr

The important property of the operator $H_2$ established in the appendix is as follows:
the Fourier coefficients $\hat h_2(\xi;\theta)$ satisfy
%\footnote{write in appendix how to make here $\tilde R$ instead of $R$} 
\bee\label{zero}
\hat h_2(\xi;\theta)=0, \ \ if\  \theta\ne 0 \ \ and \ \ \xi\not\in R(\theta). 
\ene 
This property implies that if a point $\xi$ lies outside all the resonance zones, then the one-dimensional subspace spanned by the corresponding  
$\be_{\xi}$ is invariant with respect to $H_2$. If, on the other hand, for some (unique) $\theta$ we have $\xi\in R(\theta)$, then the  two-dimensional  subspace spanned by  
$\be_{\xi}$ and $\be_{\xi+2\theta}$ is invariant with respect to $H_2$.

We have the following further properties of the symbol of $H_2$. The non-vanishing Fourier coefficients $\hat h_2(\xi;\theta)$ must have $\theta\in\Theta_{3N}$. 
Each Fourier coefficient $\hat h_2(\xi;\theta)$ is smooth outside of the end-points of $ R(\theta)$. 
Also, we have $\hat h_2(-\xi;-\theta)=\hat h_2(\xi;\theta)$ and $\hat h_2(\xi;\theta)=\overline{\hat h_2(\xi+2\theta;-\theta)}$. 
The Fourier coefficient $\hat h_2(\xi;0)$ satisfies 
\bee\label{zerocoeff}
\hat h_2(\xi;0)=|\xi|^2+\ep\tau+f(\xi;0),
\ene
where $f(\xi;0)$ is the perturbation theory correction:
\bee\label{eq:nonres1}
f(\xi;0)=\sum_{p=2}^N\ep^p f_p(\xi;0). 
\ene
%Here, each $f_p$ is a sum of the fractions of the following type. 
%The numerator of each fraction is a product of $p$ Fourier coefficients of $V$, and the denominator is a product of $p-1$ terms of the form 
%$(|\xi+2\theta_q|^2-|\xi|^2)$, $\theta_q\in\Theta$. 
Here, each $f_p$ is a sum of terms of the following type: each term is a product of $p$ Fourier coefficients of $V$ and $p-1$ functions $\tilde\chi_{\theta_j}$ (for details see Lemma~\ref{lem:symbol}). In particular,
\bee\label{eq:nonres2} 
f_2(\xi;0)=-\sum_{\theta\in\Theta'}|\hat V_{\theta}|^2\tilde\chi_{\theta}(\xi)%=
%-\sum_{\theta\in\Theta', \ \xi\not\in R(\theta)}\frac{|\hat V_{\theta}|^2}{|\xi+2\theta|^2-|\xi|^2}.
\ene
and, assuming $\xi$ is non-resonant for all $\theta$, we have 
\bee\label{eq:nonres2nr} 
f_2(\xi;0)=
-\sum_{\theta\in\Theta'}\frac{|\hat V_{\theta}|^2}{|\xi+2\theta|^2-|\xi|^2}.
\ene
Note that \eqref{zerocoeff} implies that for $\xi>0$ and sufficiently small (depending on $\xi$) $\ep$ we have
\bee\label{zeromonotone}
\frac{\partial}{\partial\xi}\hat h_2(\xi;0)>0.
\ene
%Moreover, if we have two non-resonant points $0<\xi_1<\xi_2$, then for sufficiently small $\ep$ we have $\hat h_2(\xi_1;0)<\hat h_2(\xi_2;0)$
%(this property does not necessarily follow from \eqref{zeromonotone}, since $\hat h_2(\xi;0)$ does not have to be differentiable, or even continuous on the entire interval $[\xi_1,\xi_2]$.). 

Similarly, if $\xi\in R({\theta})$ (and $\theta\ne 0$), we have 
\bee\label{eq:res}
\hat h_2(\xi;\theta)=\ep\hat V_{\theta}(1-\varphi_{\theta}(\xi))+\sum_{p=2}^N\ep^p f_p(\xi;\theta),
\ene 
where $f_p(\xi;\theta)$ has a form similar to $f_p(\xi;0)$ and, in particular,  
\bee\label{eq:nonres2a} 
f_2(\xi;\theta)=-\frac12 \sum_{\theta_1,\theta_2\in\Theta',\theta_1+\theta_2=\theta}\hat  V_{\theta_1}
\hat V_{\theta_2}
(\tilde\chi_{\theta_2}(\xi)-\tilde\chi_{\theta_2}(\xi+2\theta_1)). 
%=
%\sum_{\theta\in\theta, \ \xi\not\in R(\theta)}\frac{|\hat V_{\theta}|^2}{|%\xi+\theta|^2-|\xi|^2}.
\ene

As we have stated above, there are two types of invariant subspaces of $H_2$: 

1. If $\xi\not\in\cup_{\theta\in\Theta_{3N}} R(\theta)$, then $\be_{\xi}$ generates a one-dimensional invariant subspace of $H_2$;

2. Suppose, $\xi\in R(\theta)$ for some $\theta\in\Theta_{3N}$.
%, say $\xi\in R(\theta)^{+}$. 
Denote $\eta:=\xi+2\theta\in R(-\theta)$. Then 
the two-dimensional subspace generated by $\be_{\xi}$ and $\be_{\eta}$ is invariant under $H_2$. 

%Our definition \eqref{phi} now amounts to 
%\bee\label{phi1}
%\phi_{\theta}(\xi):=\begin{cases}
%0, & if \ \ |\xi s(\theta)-|\theta||<\ep_n^{1/2}\\
%1, & \ otherwise,
%\end{cases}
%\ene
%where we have denoted $s(\theta):=\frac{\theta}{|\theta|}$ to be the sign of $\theta$. 

\subsection{Basic Constructions}
Let us now construct the mapping $G:\R\to\R$ defined in the following way. Suppose first 
that $\xi$ is non-resonant. Then we put $G(\xi):=\hat h_2(\xi;0)=|\xi|^2+\ep\tau+f(\xi;0)$; notice that for non-resonant $\xi$ we thus have $G(-\xi)=G(\xi)$. Suppose now that $\xi$ is resonant. Then it belongs to exactly one resonance zone, say $\xi\in R(-\theta)$. Consider the $2\times 2$ matrix $M(\xi)=M_{-\theta}(\xi)$, where the diagonal elements are $\hat h_2(\xi;0)$ and $\hat h_2(\xi-2\theta;0)$ and off-diagonal elements are $\hat h_2(\xi;-\theta)$ and $\hat h_2(\xi-2\theta;\theta)$. This matrix is Hermitian and has two real eigenvalues $\la_1(M(\xi))\le \la_2(M(\xi))$. 
We define $G(\xi)=\la_2(M(\xi))$ if $\xi>0$ and 
$G(\xi)=\la_1(M(\xi))$ if $\xi<0$. Thus defined function $G$ is even outside the resonance zones, increasing for positive $\xi$ outside the resonance zones and is continuous outside the end-points of the resonance zones. At the end points of resonant zones it may have jumps. These jumps are caused by the fact that the pair of eigenvalues $(\la_1(M(\xi)), \la_2(M(\xi)))$ can be extended continuously outside a resonance zone, but once we have made a choice of one eigenvalue for each resonance point $\xi$, we have introduced discontinuities. A careful look at the situation convinces one that in fact
if we consider two resonant zones  $R(\pm\theta,\de)$, then the function $G$ may have discontinuities only  at the left  endpoints, namely $\pm\theta-\frac{\de}{4|\theta|}$, and there the jump of $G$ can be quite substantial (of order $\ep$).  At two other points 
$\pm\theta+\frac{\de}{4|\theta|}$ function $G$ is continuous. See Figure 1 for the sketch of the graph of $G$. %\footnote{picture here} 

\begin{figure}
\begin{center}
\includegraphics[width=10cm, height=10cm]{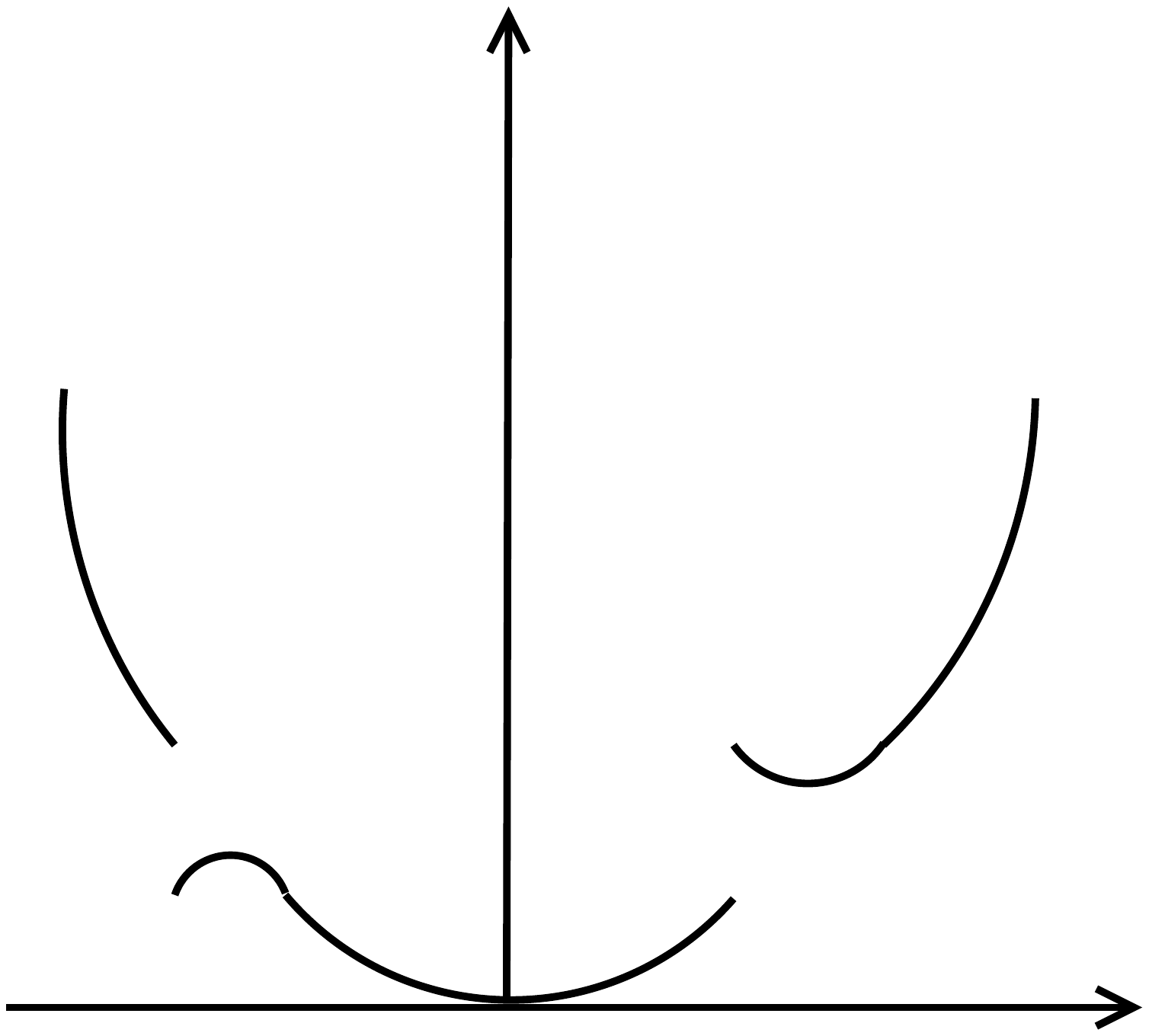}
\end{center}
\caption{The graph of G in the simplest case of one frequency}
\end{figure}

%\bel\label{jumps}
%The size of jumps of $G$ at $\pm\theta+\frac{\de}{4|\theta|}$ is $O(\ep^N)$.
%\enl
%\bep
%Consider, together with the operator $H_2$, a different operator $H_2'$ which is constructed in the same way as $H_2$, only with the parameter $\de'(N)=\de(N)/2$. Denote by $G'$ the function $G$ constructed for $H_2'$. Then function $G'$ has jumps at the endpoints of the resonance zones corresponding to $H_2'$, i.e. at points $\pm\theta+\frac{\de}{8|\theta|}$. The important property of functions $G$ and $G'$ which follows from the construction (see in particular \eqref{eq:kand}) is the following one: for any point $\xi$ 
%inside $R(\pm\theta;2\de)$ the value  $G(\xi)$ is within distance $O(\ep^N)$ to either $G'(\xi)$, or $G'(\xi\pm 2\theta)$. This, and studying Figure 2\footnote{picture here} should convince one that, since the points $\pm\theta+\frac{\de}{4|\theta|}$ are outside the resonance zones of $H_2'$, the function $G'$ must be continuous at these points and, therefore, the   size of jumps of $G$ at $\pm\theta+\frac{\de}{4|\theta|}$ is $O(\ep^N)$.
%\enp

The most important property of $G$ is the following one: we have
\bee\label{Nla}
N(\la;H_2)=(2\pi)^{-1}\meas\{\xi, \ G(\xi)\le\la\}=:(2\pi)^{-1}\meas\ \Om_{\la},
\ene
where we have denoted 
$\{\xi, \ G(\xi)\le\la\}=:\Om_{\la}=\Om_{\la}(G)$. This property was proved in \cite{PaSh1} and it immediately implies that the spectrum of $H_2$ is
\bee\label{sp}
\sigma(H_2)=\overline{\{G(\xi),\ \xi\in\R\}}.
\ene
%Notice that we have $G(\xi)=\hat h_2(\xi;0)+O(\ep)$, which implies that for $0<\xi_1<\xi_2$ and sufficiently small %$\ep$ we have $G(\xi_1)<G(\xi_2)$. 

Equation \eqref{sp} shows that in order to study the spectrum of $H_2$, we need to look at the range of $G$. Our discussions above and Figure 1 %and Lemma \ref{jumps} 
imply the following statement:
\bel
The range of $G$ consists of the entire semi-axis $[\la_0,+\infty)$, with the possible exception of the gaps, `generated' by the resonant zones. Each pair of zones $R(\pm\theta)$ generates at most one gap 
$(a_-(\theta),a_+(\theta))$. 
%of size bigger than $O(\ep^N)$, where 
%$a_{\pm}(\theta)=\theta^2+O(\delta)$. There may be another `artificial' gaps (one or two for each pair of resonance zones) generated by the jumps of $G$ at the endpoints  
%$\pm\theta+\frac{\de}{4|\theta|}$, but their size is $O(\ep^N)$ (and they will decrease to zero when we increase $N$). 
\enl
%\ber
%The existence of such artificial gaps is a side-effect of considering cut-off functions \eqref{phi} with jumps. We could have considered continuous cut-off functions instead; this would have eliminated artificial gaps, but would have led to extra difficulties elsewhere. 
%\enr

%\bep
%This follows from the definition of $G$ and the following obvious property: suppose, $A$ is a Hermitian matrix with diagonal elements $a_1<a_2$. Then the smallest eigenvalues of $A$ is at most $a_1$, and the biggest one is at least $a_2$. 
%\enp
Later, we will obtain more precise information on the location and the length of the gaps. 

 Let us introduce a change of variables: we put $\xi=\theta_0+\zeta\in R(-\theta_0)$, so that $|\zeta|\le\frac{{\delta}}{4|\theta_0|}$. We also put 
\bee\label{notation}
\bes
&s=s(\zeta):=\frac{f(\zeta-\theta_0;0)+f(\zeta+\theta_0;0)}{2}=O(\ep^2), \ \\
&t=t(\zeta):=\frac{f(\zeta-\theta_0;0)-f(\zeta+\theta_0;0)}{2}=O(\ep^2), \ \\
&\nu:=\hat V(\theta_0), \ \\
&g=g(\zeta):=\hat h_2(\zeta+\theta_0;-\theta_0)-\nu\ep=O(\ep^2), \\
&s_j=s_j(\zeta):=\frac{f_j(\zeta-\theta_0;0)+f_j(\zeta+\theta_0;0)}{2},\\
&t_j=t_j(\zeta):=\frac{f_j(\zeta-\theta_0;0)-f_j(\zeta+\theta_0;0)}{2},\\
&g_j=g_j(\zeta):=f_j(\zeta+\theta_0;-\theta_0).
\end{split}
\ene
%Finally, we denote $\mu:=\ep\tau$. 
%Note that since we are considering the case $\theta_0\in\theta$, we have $\nu\ne 0$. 

The diagonal elements of $M(\xi)$ are equal to $(\zeta-\theta_0)^2+(s+t)+\mu$ and 
$(\zeta+\theta_0)^2+(s-t)+\mu$, $\mu:=\varepsilon\tau$ and the off-diagonal elements are $\ep\nu+g$ and $\ep\bar\nu+\bar g$. 

The characteristic polynomial %(with variable $\sigma$) 
of the matrix $M(\xi)-\mu$ is
\bee
\bes
&\chi(\sigma)=\sigma^2-2\sigma[(\zeta^2+\theta_0^2)+s]\\
&+[(\zeta^2-\theta_0^2)^2+2s(\zeta^2+\theta_0^2)-|\nu^2|\ep^2+4t\zeta\theta_0+(s^2-t^2-2\ep\Re (\nu\bar g)-|g|^2)]
\end{split}
\ene
and the eigenvalues of $M(\xi)$ are 
\bee\label{sigma0}
\sigma_{\pm}=\sigma_{\pm}(\zeta;\theta_0;\mu)=\zeta^2+\theta_0^2+\mu+s\pm ((2\zeta\theta_0-t)^2+|\varepsilon\nu+g|^2)^{1/2}. 
\ene
Obviously, $\la_2(M(\xi))=\sigma_{+}(\zeta)$, and $\la_1(M(\xi))=\sigma_{-}(\zeta)$. 

\subsection{Spectral gaps}
Let us find the size of the spectral gap around $\theta_0^2+\mu$. This gap is an interval $[\sigma_-^{\max},\sigma_+^{\min}]$, where $\sigma_-^{\max}$ is the maximal value of $\sigma_{-}(\zeta;\theta_0;\mu)$ when $\zeta$ runs over the interval $[-\frac{{\delta}}{4|\theta_0|},\frac{{\delta}}{4|\theta_0|}]$, and $\sigma_+^{\min}$ is the minimal value of $\sigma_{+}(\zeta;\theta_0;\mu)$. 

\ber%\footnote{one} 
It is easy to see and will be even clearer in what follows that all the objects we are interested in require detailed information only from the interior of the resonant zones. In particular, maximum value of $\sigma_-$ and minimal value of $\sigma_+$ are attained inside the interval $[-\frac{{\delta}}{100|\theta_0|},\frac{{\delta}}{100|\theta_0|}]$ (assuming of course, $\varepsilon$ is small enough). This allows us to ignore cut-off functions $\varphi_\theta$ introduced above as they are equal to zero in the region of interest.
\enr

Recall that $|\mu|\ll\ep$ and $|s|,\ |t|, \ |g|\ll \ep^2$. Moreover, it follows from the definitions and properties of $\hat h_2$ that $t(0)=s'(0)=0$. Thus, we can rewrite
\bee
\sigma_{\pm}=\sigma_{\pm}(\zeta;\theta_0;\mu)=\zeta^2(1+O(\varepsilon^2))+\theta_0^2+\mu+s(0)\pm (4\zeta^2\theta_0^2(1+O(\varepsilon^2))+|\varepsilon\nu+g|^2)^{1/2}. 
\ene

If $\hat h_2(\theta_0;-\theta_0)=0$, which in particular means that $\nu=g(0)=0$ then, obviously, $\sigma_+(0)=\sigma_-(0)=\theta_0^2+\mu+s(0)$ and thus, we have no gap. Assume now that $\hat h_2(\theta_0;-\theta_0)\not=0$. Let us use the analytic expansion for 
$$|\hat h_2(\theta_0+\zeta;-\theta_0)|^2=|\varepsilon\nu+g|^2$$ 
in powers of $\zeta$ in the neighbourhood of $\zeta=0$ as $(\varepsilon\nu+g(\zeta))\overline{(\varepsilon\nu+g(\bar\zeta))}$.
\ber
This complete expansion exists (both here and below) because all cut-off functions $\varphi_\theta$ are equal either to 0 or 1 in the regions under consideration. For more details see Appendix, in particular, Remark~\ref{quasiperiodic} and Lemma~\ref{lem:symbol}
\enr

Thus, let us denote by $2p$, $p\in\N$ the first non-trivial power in the expansion of $|\hat h_2(\theta_0;-\theta_0)|^2$ in powers of $\ep$; obviously, this power has to be even since the expression is non-negative. We put $c_1(\varepsilon):=\varepsilon^{-2p}|\hat h_2(\theta_0;-\theta_0)|^2$, $c_1(0)>0$. We then have 
\bee\label{analyt}
|\varepsilon\nu+g|^2=c_1(\varepsilon)\varepsilon^{2p}+c_2(\varepsilon)\varepsilon^{p+q}\zeta+\zeta^2O(\varepsilon^3),\ \ \ c_1(0)>0, \ \ \ q\ge p,
\ene
where $c_2(\varepsilon)$ is analytic in $\varepsilon$ and either $q\in\N$ (then it is easy to see that $q\ge 2$), or $q=\infty$ (meaning $c_2\equiv0$). 
In particular, we have $p=1$ and $c_1(0)=|\nu|^2$ for $\nu\not=0$, or $p\geq2$ for $\nu=0$. %We also have $q\geq2$. 
Suppose first that $q=+\infty$ (i.e. $c_2\equiv 0$). Then,  obviously, we have $\sigma_{-}^{\max}=\sigma_{-}(0)$ and $\sigma_{+}^{\min}=\sigma_{+}(0)$, which implies that  
the size of the gap is exactly $2|\hat h_2(\theta_0,-\theta_0)|=2\sqrt{c_1\varepsilon^{2p}}$. Thus, we can assume that $q$ is finite. Calculating derivatives, we obtain the following equations for any critical points $\zeta_\pm$ of $\sigma_\pm$:
\bee\label{sigma1}
2\zeta(1+O(\varepsilon^2))(4\zeta^2\theta_0^2(1+O(\varepsilon^2))+|\varepsilon\nu+g|^2)^{1/2}\pm(4\zeta\theta_0^2(1+O(\varepsilon^2))+c_2\varepsilon^{p+q})=0.
\ene
We immediately see that $\zeta_\pm=O(\varepsilon^{p+q})$. %(the case $q=\infty$ or $c_2=0$ corresponds to $\zeta_\pm=0$). 
Then we can rescale $\zeta=:\varepsilon^{p+q}z$ and rewrite \eqref{sigma1} as follows (here we also use \eqref{analyt}):
\bee\label{sigma2}
2z(1+O(\varepsilon^2))\varepsilon^p(c_1+O(\varepsilon^{2q}))^{1/2}\pm(4z\theta_0^2(1+O(\varepsilon^2))+c_2)=0.
\ene
This shows that the solutions $z_\pm$ and thus $\zeta_\pm$ are analytic in $\varepsilon$ with the main term $\zeta_\pm\sim-\frac{c_2\varepsilon^{p+q}}{4\theta_0^2}$. Plugging this %(or $\zeta_\pm=0$ for $q=\infty$ ($c_2=0$)) 
into \eqref{sigma0} and using the fact that $\sigma_-^{\max}=\sigma_-(\zeta_-)$ and $\sigma_+^{\min}=\sigma_+(\zeta_+)$, we obtain the following lemma

\begin{lem}\label{gaps}
If $\hat h_2(\theta_0;-\theta_0)\equiv 0$ (as a function of $\varepsilon$), then the size of the gap is $O(\varepsilon^N)$. If $\hat h_2(\theta_0;-\theta_0)\not\equiv 0$, then there exists the complete asymptotic expansion for the size of the gap (up to the order $O(\varepsilon^N)$) starting with twice the first non-zero power of $\varepsilon$ in expansion \eqref{eq:res} of $|\hat h_2(\theta_0;-\theta_0)|$, i.e. either $2|\nu|\varepsilon$ or $2|f_p(\theta_0,-\theta_0)|\varepsilon^p$, $p\geq2$.%\footnote{two}
 %\footnote{let us write the precise expression for the first term, with the coefficient}
%\footnote{Since in our procedure we update $h_2$ with every new step, it actually means that for all gaps we have complete asymptotic expansion, while some expansions can lead to super power behaviour or be closed.}\footnote{Can we get a superpower, but not closed gap?}
\end{lem}

Since $N$ is arbitrary, we immediately obtain the following result (which also implies Theorem~\ref{gapst1int} for quasi-periodic case):

\bet \label{gapst} Suppose, $V$ is quasi-periodic. 
If $\hat h_2(\theta_0;-\theta_0)\equiv 0$ for each $N$, then the size of the spectral gap around $\theta_0^2$ is $O(\varepsilon^{\infty})$. If $\hat h_2(\theta_0;-\theta_0)\not\equiv 0$ for some $N$, then there exists the complete asymptotic expansion for the size of the gap in natural powers of $\ep$; the first term of this expansion is twice the first non-zero term in expansion \eqref{eq:res} of $|\hat h_2(\theta_0;-\theta_0)|$, i.e. either $2|\nu|\varepsilon$ or $2|f_p(\theta_0,-\theta_0)|\varepsilon^p$, $p\geq2$.
\ent
\ber
1) As one can see from the proof, we also have complete expansions of the upper and lower ends of each gap. 

2) We notice that while $\hat{h}_2$ depends on the cut-off function $\varphi$, $\hat{h}_2(\theta_0,-\theta_0)$ does not. Thus, as it should be, corresponding expansion from Theorem~\ref{gapst} is independent of the particular choice of the cut-off function.

\enr

\subsection{Integrated Density of States}
Now let us discuss the IDS of $H_2$. 
Formula \eqref{Nla} implies that in order to study the integrated density of states, we need to solve the equation 
\bee\label{equation}
G(\xi)=\la. 
\ene
In the unperturbed case (when $G(\xi)=\xi^2$) this equation has two solutions whenever $\la>0$. After the perturbation, this equation may have no solutions (when $\la$ is inside a spectral gap), or it may have one solution (when $\la$ is exactly at the spectral edge of $H_2$). As we will see later, in other cases equation \eqref{equation} has exactly two solutions. 

%, but also sometimes the number of solutions may be bigger than two. Outside of the interior of resonance zones this could happen when function $G$ has a `negative' jump at the right endpoints of the resonance zones  $\pm\theta+\frac{\de}{4|\theta|}$ (see Figure 3). Lemma \ref{jumps}, \eqref{eq:kand}, and further discussion in this subsection imply that when such a negative jump occurs, the distance between two solutions generated by such negative `jump' is $O(\ep^N)$. Since we want to compute \eqref{Nla} only up to $O(\ep^N)$, we will not distinguish such extra `artificial' solutions of the equation $G(\xi)=\la$; after making this convention, we can safely state that the number of solutions of equation $G(\xi)=\la$ is never bigger than two for $\sqrt\lambda$ being outside of the interior of the resonance zones. %\footnote{update abzats; I am not sure now we actually prove there are no more than 2 solutions, since we give asymptotics of the gap edges in $\epsilon$ but not in $\xi$. We probably can just refer to effective masses results...}

If $\la$ is negative, the above constructions imply that $N(\la;H)=0$ for
sufficiently small $\ep$.
Suppose now that $\la$ is positive and $\sqrt{\la}\not\in\Theta_{3N}$ (in particular, $\la\ne 0$).
%If $\la$ is negative, the above constructions imply that $N(\la;H)=0$ for
%sufficiently small $\ep$. Suppose that $\la$ is positive.  
Then, for sufficiently small $\de$, points (both of them) $\xi$ with $\xi^2=\la$ do not belong to any resonance region; the same is true for points of the form $\la-\ep\tau$. %\footnote{either explain from diophantine property, or use finiteness of the number of resonance zones} 
This, together with \eqref{zeromonotone}, implies that the equation $G(\xi)=\la$ has two solutions (recall that we use convention of not distinguishing two solutions that are within distance $O(\ep^N)$ from each other), call them $G^{-1}(\la)>0$ and $-G^{-1}(\la)$. Monotonicity of $G$ implies that (again for sufficiently small $\de$) the following holds: whenever $0<\eta<G^{-1}(\la)$, we have $G(\eta)<\la$, and whenever $\eta>G^{-1}(\la)$, we have $G(\eta)>\la$. 

This implies that $\Om_{\la}=[-G^{-1}(\la),G^{-1}(\la)]$, so that 
\bee
N(\la;H_2)=\pi^{-1}G^{-1}(\la).
\ene
Now an easy application of the inverse function theorem and \eqref{eq:nonres1}-\eqref{eq:nonres2} imply
\bee\label{nonresas0}
N(\la;H_2)=\pi^{-1}\sqrt\la+\sum_{p=1}^N a_{p}(\la) \ep^p+O(\ep^N).
\ene
Since $N$ is arbitrary, formulas \eqref{N} and \eqref{eq:N1} now imply 
\bee\label{nonresas}
N(\la;H)\sim \pi^{-1}\sqrt\la+\sum_{p=1}^{\infty} a_{p}(\la) \ep^p.
\ene

The next case we consider is $\la=\theta_0^2>0$ with either  $\theta_0\in\Theta$ (so that $\hat V_\theta\ne 0$), or $\theta_0\in\Theta_m$ for some $m>1$;  
for the sake of definiteness we will also assume that $\theta_0>0$. Then we have the following:
\bee\label{eq:Om}
\bes
&
\Om_{\la}=\{\xi, \ |\xi|^2<\la, \xi\not\in (R(\theta_0)\cap R(-\theta_0))\}\\
&
\sqcup
\{\xi\in R(-\theta_0), \  \la_2(M(\xi))\le\la\}\sqcup
\{\xi\in R(\theta_0), \  \la_1(M(\xi))\le\la\}
\end{split}
\ene
(the disjoint union) and, therefore,
\bee\label{eq:Om3b}
\bes
&(2\pi)N(\la;H_2)=2\sqrt\la-\frac{{\delta}}{2|\theta_0|}+
\meas\{\xi\in R(-\theta_0), \  \la_2(M(\xi))\le\la\}\\
&+
\meas\{\xi\in R(\theta_0), \  \la_1(M(\xi))\le\la\}.
\end{split}
\ene
%where $\meas$ stands for the Lebesgue measure. 
Note that 
\bees
\meas\{\xi\in R(\theta_0), \  \la_1(M(\xi))\le\la\}=\meas\{\xi\in R(-\theta_0), \  \la_1(M(\xi))\le\la\},
\enes
and so
\bee\label{eq:Om3}
\bes
&
(2\pi)N(\la;H_2)=2\sqrt\la-\frac{{\delta}}{2|\theta_0|}+
\meas\{\xi\in R(-\theta_0), \  \la_2(M(\xi))\le\la\}\\
&+
\meas\{\xi\in R(-\theta_0), \  \la_1(M(\xi))\le\la\}.
\end{split}
\ene
%Let us define two new functions $g^1_{\theta}(\xi)$ and $g^2_{\theta}(\xi)$ in the following way: Given arbitrary point $\xi\in\R$ (not necessarily in the resonant zone $R(\theta)$), we construct the matrix $M_{\theta}(\xi)$ in the same way as before (the diagonal elements are $\hat h_2(\xi;0)$ and $\hat h_2(\xi+\theta;0)$ and off-diagonal elements are $\hat h_2(\xi;\theta)$ and $\hat h_2(\xi+\theta;-\theta)$). This matrix has two real eigenvalues $\la_1(M_{\theta}(\xi))\le \la_2(M_{\theta}(\xi))$, and we put $g^j_{\theta}(\xi)=\la_j(M_{\theta}(\xi))$, $j=1,2$. It is easy to check that for small $\ep$ the following holds: suppose $\boldeta\not\in (R(\theta)\cap R(-\theta))$. Then we always have $g^2_{\theta}(\xi)>\la$, and the inequality $g^1_{\theta}(\xi)<\la$ is equivalent to $|\boldeta|<\theta$, and $g^2_{\theta}(\xi)<\la$ is equivalent to $|\boldeta|<\theta$. This together with \eqref{eq:Om} implies that 
%\bee\label{eq:Om1}
%\Om_{\la}(g)=\Om_{\la}(g_{\theta}).
%\ene
%Moreover, an easy check convinces us that \eqref{eq:Om3} still holds if we replace $\la=\theta^2$ by $\la=\theta^2+\mu$, where $\mu\ll\ep$; we will need this expansion of the range of $\la$ when we will consider potentials with non-zero average, so most often we will assume $\mu=\tau\ep$. 
Thus, we  have to study the behaviour of the eigenvalues of $M(\xi)$ when $\xi\in R(-\theta_0)$.

As we have mentioned above, in order to calculate it, we need to compute the measure of the set of points $\zeta$, $|\zeta|\le\frac{{\delta}}{4|\theta_0|}$ for which 
\bee
\zeta^2+(s+\mu)\pm (4\zeta^2\theta_0^2+|\nu|^2\ep^2-4t\zeta\theta_0+t^2 +2\ep\Re (\nu\bar g)+|g|^2)^{1/2}\le 0.
\ene
%First, we will deal with the (generic) case\footnote{maybe we need to assume this on a later stage; discuss what to do if $s_2(0)\tau+\Re (\nu\bar g_2(0))=0$}  

%\bee\label{eq:s2}
%s_2(0)\ne 0. 
%\ene
%\bee\label{eq:s2}
%s_2(0)\tau+\Re (\nu\bar g_2(0))\ne 0
%\ene

We start by assuming that $\nu\ne 0$, i.e. $\theta\in\Theta$. 

Case 1. 
Assume first that $\mu\ge 0$. Then we always have  
\bee
\sigma_{+}-\theta_0^2=\zeta^2+(s+\mu)+ (4\zeta^2\theta_0^2+|\nu|^2\ep^2-4t\zeta\theta_0+t^2 +2\ep\Re (\nu\bar g)+|g|^2)^{1/2}> 0
\ene
and, thus, 
\bee
\meas\{\xi\in R(-\theta_0), \  \la_2(M(\xi))\le\la\}=0.
\ene
Therefore, we need to consider only the measure of $\zeta$, $|\zeta|\le\frac{{\delta}}{4|\theta_0|}$ for which  
\bee\label{eq:1}
\sigma_{-}-\theta_0^2=\zeta^2+(s+\mu)- (4\zeta^2\theta_0^2+|\nu|^2\ep^2+2\ep\Re (\nu\bar g)+\tilde g)^{1/2}\le 0, 
\ene
where we have denoted
\bee
\tilde g=\tilde g(\zeta):=-4t\zeta\theta_0+t^2 +|g|^2.
\ene
Note that $\tilde g=O(\ep^{4})+\zeta^2 O(\ep^2)$. 
Inequality \eqref{eq:1} is equivalent to  
\bee\label{eq:2}
\zeta^4+2(s+\mu)\zeta^2+(s+\mu)^2 \le 4\zeta^2\theta_0^2+|\nu|^2\ep^2+2\ep\Re (\nu\bar g)+\tilde g, 
\ene
or
\bee\label{eq:3}
\zeta^4-2[2\theta_0^2-(s+\mu)]\zeta^2+[(s+\mu)^2 -|\nu|^2\ep^2-2\ep\Re (\nu\bar g)-\tilde g]\le 0.
\ene
Let us formally solve \eqref{eq:3} as a quadratic equation in $\zeta^2$. We obtain:
\bee\label{eq:4}
\zeta^2_{\pm}=[2\theta_0^2-(s+\mu)]\pm \sqrt{[2\theta_0^2-(s+\mu)]^2-[(s+\mu)^2 -|\nu|^2\ep^2-2\ep\Re (\nu\bar g)-\tilde g]}.  
\ene
Expanding the square root in the RHS of \eqref{eq:4}, we obtain 
\bee\label{eq:4p}
\zeta^2_{+}=2[2\theta_0^2-(s+\mu)]+
\sum_{j=1}^\infty (-1)^j {{j}\choose{1/2}}
[2\theta_0^2-(s+\mu)]^{-2j+1} [(s+\mu)^2 -|\nu|^2\ep^2-2\ep\Re (\nu\bar g)-\tilde g]^j 
\ene
and
\bee\label{eq:4m}
\zeta^2_{-}=-
\sum_{j=1}^\infty (-1)^j {{j}\choose{1/2}}
[2\theta_0^2-(s+\mu)]^{-2j+1} [(s+\mu)^2 -|\nu|^2\ep^2-2\ep\Re (\nu\bar g)-\tilde g]^j 
\ene

A straightforward application of the inverse function theorem implies that equation \eqref{eq:4p} (where the functions $s$ and $\tilde g$ have $\zeta_{+}$ as their argument) has exactly one positive solution $\zeta_{+}^{+}=4\theta_0^2+O(\ep)$ and one negative solution
$\zeta_{+}^{-}=-4\theta_0^2+O(\ep)$. 
Therefore, all points $\zeta\in [-\frac{{\delta}}{4|\theta_0|},\frac{{\delta}}{4|\theta_0|}]$  satisfy the inequality $\zeta^2<\zeta_{+}^2$.   

The situation with $\zeta_{-}$ is a bit more involved.  
There are two further sub-cases. Suppose first that for all $\zeta\in [-\frac{{\delta}}{4|\theta_0|},\frac{{\delta}}{4|\theta_0|}]$ we have $[(s+\mu)^2 -|\nu|^2\ep^2-2\ep\Re (\nu\bar g)-\tilde g]<0$ (this happens, for example, if $\mu=\tau\ep$ with $|\tau|<|\nu|$). Then the RHS of equation \eqref{eq:4m} is negative, so it has no real solutions. Thus, \eqref{eq:1} holds for all $\zeta\in [-\frac{{\delta}}{4|\theta_0|},\frac{{\delta}}{4|\theta_0|}]$, and therefore in this case 
\bee
\meas\{\xi\in R(-\theta_0), \  \la_1(M(\xi))\le\la\}=\frac{{\delta}}{2|\theta_0|}
\ene
and 
\bee\label{eq:Nconst}
N(\theta_0^2;H_2)=\pi^{-1}\theta_0 
\ene
for all small $\ep$. The situation is the same when $|\tau|=|\nu|$ and $[s_2(0)\tau-\Re (\nu\bar g_2(0))]<0$.%, i.e.  $s_2(0)<0$

Now assume that $\mu=\tau\ep$ with $|\tau|>|\nu|$. Then the inverse function theorem implies that \eqref{eq:4m} has a unique positive and a unique negative solutions $\zeta_{-}^{\pm}\sim\pm\ep$; both have a complete asymptotic expansion in powers of $\ep$. In this case we have 
\bee
\meas\{\xi\in R(-\theta_0), \  \la_1(M(\xi))\le\la\}=\frac{{\delta}}{2|\theta_0|}+\sum_{p=1}^N\tilde  a_p \ep^p+O(\ep^N).
\ene 
and
\bee\label{eq:N2}
N(\theta_0^2;H_2)= \pi^{-1}\theta_0+\sum_{p=1}^N a_p \ep^p+O(\ep^N),\ \ a_1<0.
\ene
%It is easy to check that the situation is the same if 
%$\mu=\nu=0$ and the following (generic) condition is satisfied:

%\bee\label{eq:s2}
%s_2(0)\tau+\Re (\nu\bar g_2(0))\ne 0
%\ene

Finally, let us consider the most interesting situation when $\mu=\tau\ep$ with $|\tau|=|\nu|$ and $ [s_2(0)\tau-\Re (\nu\bar g_2(0))]>0$. % (i.e.  $s_2(0)>0$). 
In this case, equation  \eqref{eq:4m} has a unique positive solution and a unique negative solution $\zeta_{-}^{\pm}\sim \pm\ep^{3/2}$; both  $\zeta_{-}^{\pm}\ep^{-3/2}$ have complete asymptotic expansions in powers of $\ep$.  
In this case we have 
\bee
\meas\{\xi\in R(-\theta_0), \  \la_1(M(\xi))\le\la\}=\frac{{\delta}}{2|\theta_0|}+\ep^{3/2}\sum_{p=0}^N\tilde  a_p \ep^p+O(\ep^N).
\ene 
and
\bee\label{eq:N3/2}
N(\theta_0^2;H_2)= \pi^{-1}\theta_0+\ep^{3/2}\sum_{p=0}^{N-2} a_p \ep^p+O(\ep^N),\ \ a_0<0.
\ene
%Note that in this case the corresponding expansions are valid only for one choice of the sign of $\ep$, while for the %other choice the expansion is given by \eqref{eq:Nconst}.

Case 2. 
Now we discuss the case $\mu<0$. In this case, as we have already mentioned, we always have
\bee
\zeta^2+(s+\mu)- (4\zeta^2\theta_0^2+|\nu|^2\ep^2-4t\zeta\theta_0+t^2 +2\ep\Re (\nu\bar g)+|g|^2)^{1/2}\le 0
\ene
whenever $|\zeta|<\frac{\delta}{4|\theta_0|}$ 
and, thus, 
\bee
\meas\{\xi\in R(-\theta_0), \  \la_1(M(\xi))\le\la\}=\frac{{\delta}}{2|\theta_0|}.
\ene
 Therefore, we only need to solve the inequality 
\bee
\zeta^2+(s+\mu)+ (4\zeta^2\theta_0^2+|\nu|^2\ep^2-4t\zeta\theta_0+t^2 +2\ep\Re (\nu\bar g)+|g|^2)^{1/2}> 0.
\ene
Calculations similar to those above imply the following.

If $|\tau|<|\nu|$, or if $|\tau|=|\nu|$ and $[s_2(0)\tau-\Re (\nu\bar g_2(0))]<0$, we have 
\bee\label{eq:Nconst1a}
N(\theta_0^2;H_2)=\pi^{-1}\theta_0. 
\ene

If $|\tau|>|\nu|$,  %(i.e.  $s_2(0)>0$), 
we have 
\bee\label{eq:Nconst1b}
N(\theta_0^2;H_2)= \pi^{-1}\theta_0+\sum_{p=1}^N a_p \ep^p+o(\ep^N),\ \ a_1<0.
\ene 

Finally, if $|\tau|=|\nu|$ and $\ep[s_2(0)\tau-\Re (\nu\bar g_2(0))]>0$,  %(i.e.  $s_2(0)<0$), 
we have 
\bee\label{eq:N3/2a}
N(\theta_0^2;H_2)\sim \pi^{-1}\theta_0+\ep^{3/2}\sum_{p=0}^{N-2} a_p \ep^p+o(\ep^N),\ \ a_0<0.
\ene

The situation when $|\tau|=|\nu|$ and $[s_2(0)\tau-\Re (\nu\bar g_2(0))]=0$ (in either case 1 or 2) can be dealt with similarly. It is not hard to see that $N(\theta_0^2;H_2)$ admits the asymptotic expansion of the form 
\bee\label{eq:N3/2anew}
N(\theta_0^2;H_2)\sim \pi^{-1}\theta_0+\ep^{k/2}\sum_{p=0}^{N-2} a_p \ep^p+o(\ep^N),\ \ a_0<0,
\ene
with some natural $k\geq4$ including the case $k=\infty$. We omit the details of this calculation.

Suppose now that $\nu=0$, so $\theta_0\in\Theta_m\setminus\Theta_{m-1}$ with $m>1$. Then we immediately have $g_j\equiv 0$ for all $j<m$. 
If $\tau\ne 0$, then we effectively have the situation discussed above, so the results will be the same (i.e. formula \eqref{eq:Nconst1b} holds). Assume that $\tau$ also vanishes. 
Then we have
\bee
\sigma_{\pm}-\theta_0^2=\zeta^2+s\pm (4\zeta^2\theta_0^2-4t\zeta\theta_0+t^2 +|g|^2)^{1/2}=\zeta^2+s\pm ((2\zeta\theta_0-t)^2 +|g|^2)^{1/2}. 
\ene
Since $t(0)=s'(0)=0$, it is not difficult to see that (depending on the sign of $s(0)$) we either always have $\sigma_+-\theta_0^2>0$ or $\sigma_--\theta_0^2<0$. Therefore, we can repeat the above constructions. In this situation the new generic assumption will be $s_2(0)^2\not=|g_2(0)|^2$ (which for $m\ge 3$ simply means $s_2(0)\ne 0$). 
Then the calculations similar to those above imply that%\footnote{don't we have the modulus in the RHS?} 

\bee\label{eq:Nconst1c}
N(\theta_0^2;H_2)= \pi^{-1}\theta_0+\sum_{p=2}^N a_p \ep^p+o(\ep^N),\ \ a_2<0,
\ene 
for $s_2(0)^2>|g_2(0)|^2$ and \eqref{eq:Nconst1a} holds for $s_2(0)^2<|g_2(0)|^2$.

As above, the case $\nu=\tau=s_2(0)^2-|g_2(0)|^2=0$ can be considered in the same way and leads to the asymptotics \eqref{eq:N3/2anew} with possible value of $k\geq5$. %Equations \eqref{N} and \eqref{eq:N1} now lead to the following Theorem:
%We summarise the results we have proved in the following Theorem:

The last case we have to consider is $\la=0$. The only points $\xi$ where there is a chance that $G(\xi)$ is negative are located in a $(1+|\tau|)^{1/2}\ep^{1/2}$-neighbourhood of the origin and are not located in any resonance zone. Therefore, we have 

\bee
(2\pi)N(0;H_2)=\meas\{\xi,\ \hat h_2(\xi;0)<0\}=\meas\{\xi,\ |\xi|^2+\ep\tau+\sum_{p=2}^N\ep^p f_p(\xi;0)+O(\ep^N)<0\}.
\ene

Now the simple use of the Implicit Function Theorem immediately gives the answer. If $\tau>0$, then $N(0;H_2)=0$ for small $\ep$. If $\tau<0$, then 
\bee\label{lambda=0}
N(0;H_2)\sim \ep^{1/2}(\pi^{-1}|\tau|^{1/2}+\sum_{p=1}^{N-1} a_p \ep^p)+O(\ep^N).
\ene
Finally, if $\tau=0$, then we have to note that formula \eqref{eq:nonres2} implies that for small $\xi$ and non-trivial $V$ we have $f_2(\xi;0)<0$ and, therefore, 
\bee\label{lambda=01}
N(0;H_2)\sim \ep\sum_{p=0}^{N-1} a_p \ep^p+O(\ep^N),\ \ a_0>0.
\ene
 
All the asymptotic formulas for $N(\la;H_2)$ obtained above together with 
equations \eqref{N} and \eqref{eq:N1} lead to Theorem~\ref{IDS}. Again, it is easy to see that the corresponding expansions are independent of the particular choice of the cut-off function $\varphi$.

\section{Almost-periodic potential}

Let us discuss the situation when the potential is not quasi-periodic, but smooth almost-periodic, i.e. $\Theta$ is still a finite set, but we have 

\bee
V=\sum_{\theta\in\Theta_{\infty}}\hat V_{\theta}\be_{2\theta}
\ene
with 
\bee
|\hat V_{\theta}|\ll m^{-P}
\ene
for $\theta\in T_m$ and arbitrary positive $P$; recall that
\bee
T_m:=\Theta_m\setminus\Theta_{m-1}. 
\ene
We also assume that $\Theta$ satisfies the diophantine condition, i.e. for $\theta\in \Theta_m$ we have $|\theta|\gg m^{-P_0}$, where $P_0>0$ is fixed. 
\ber
We can relax the diophantine properties of the frequencies if we assume a faster decay of the Fourier coefficients: the only condition that we effectively need is that the resonance zones do not intersect, see \eqref{dioph}. 
\enr

The way we perform the gauge transform is, essentially, the same as in the quasi-periodic case, with one important difference: we cannot afford to have infinitely many resonance zones, therefore, before transforming the operator $H$ to $H_1$ and $H_2$ as above, we need to turn $H$ to a quasi-periodic operator by truncating the potential $V$. The level of the truncation depends on the size of $\ep$ -- the smaller $\ep$, the more frequencies (and resonance zones) we need to keep. Thus,  the number of resonance zones will be finite for each fixed $\ep$, but, as opposed to the quasi-periodic situation, will increase as $\ep$ goes to zero. More specifically, let us assume first that $0<\ep<\ep_0$, where $\ep_0$ is a positive number, to be chosen later. We put 
$\ep_n:=2^{-n}\ep_0$ and $I_n:=[\frac{\ep_n}{4},\ep_n]$. 
The gauge transform construction will be performed separately for each $I_n$ and the asymptotic expansions we will obtain will hold only for $\ep\in I_n$. In order to `glue' these expansions together at the end, we will use the following lemma:

\bel\label{main_lem}
Let $\alpha>0$ and let $F=F(\varepsilon)$ be a complex-valued  function defined on $(0,\ep_0)$. Suppose that for any natural numbers $M$ and $n$ there exists $M_1=M_1(M,n)\in\N$ 
%and complex numbers $a_j, j=1,...,M_1$ 
such that for $\ep\in I_n$ we have:
\bee\label{eq:main_lem1}
F(\varepsilon)=
\sum_{j=1}^{ M_1} a_{j;n}\ep^{\al j}
+O(\ep_n^{M}).
\ene
Here, $a_{j;n}$ are some coefficients depending on $j$ and $n$ (and $M$) satisfying 
\bee\label{eq:main_lem2}
a_{j;n}=O(\ep_n^{-(2\al j/3)-100})%\ \ c<4/6.
\ene
and the constants in the $O$-terms do not depend on $n$ (but they may depend on $M$). %The value of $a$ does not depend on either $n$ or $M$. 
Then there exist complex numbers $\{a_j\}$, $j=1,...,[\frac{M}{\alpha}]+1$ such that  for all $\ep$, $0<\ep<\ep_0$ we have:
\bee\label{eq:main_lem3}
F(\varepsilon)=
\sum_{j=1}^{ [\frac{M}{\alpha}]+1} a_{j}\ep^{\al j}
+O(\ep^{M}).
\ene
\enl
%We also fix $N\in\N$.

This Lemma (in slightly different form) is proved in Section 3 of \cite{PaSh1}; see also \cite{MPSh}.  In order to apply it, we have to establish \eqref{eq:main_lem1}-\eqref{eq:main_lem2}. Whenever we will be using this lemma, it will be rather straightforward to check estimates \eqref{eq:main_lem2} for the coefficients from the constructions, so in what follows we will concentrate on establishing  \eqref{eq:main_lem1}. 

\ber\label{rem:new1}
Note that \eqref{eq:main_lem1} is not a `proper' asymptotic formula, since the coefficients
$a_{j;n}$ are allowed to grow with $n$.
\enr

Now, we will describe the construction in more detail. 
Let us fix a natural number $N$ (which signifies that our errors are going to be $O(\ep^N)$) and suppose that $\ep\in I_n$. All the constructions below depend on the choice of $(n,N)$, but we will often omit writing $n$ and $N$ as the variables. Recall that for each $\theta\in\Theta'_{\infty}$ we define $Z(\theta):=m$ for $\theta\in T_m$. 
We also fix the smoothness $P$ of the potential so that 
\bee\label{P}
|\hat V_{\theta}|\ll Z(\theta)^{-P}; 
\ene
this (large) $P$ depends on $P_0$ and $N$ and will be chosen later.  
For each natural $L$ we define the truncated potential  
\bee
V_L:=\sum_{\theta\in\Theta_{L}}\hat V_{\theta}\be_{2\theta}. 
\ene
Estimate \eqref{number} implies  
\bee\label{LL}
||V-V_L||_{\infty}\ll\sum_{Z>L}Z^{-P}\#(T_Z)\ll \sum_{Z>L}Z^{3l-P}\ll
 L^{3l+1-P}<L^{-P/2}, 
\ene
assuming of course that $P$ is sufficiently large. 
Now, we choose $\tilde L=\tilde L(n,N)$ so large that the norm of the operator of multiplication by $V-V_L$ is smaller than $\ep_n^N$. The previous estimate shows that it is enough to take 
\bee\label{tildeL}
\tilde L(n;N):=\ep_n^{-\frac{2N}{P}} 
\ene
to achieve this. Then we run $3N$ steps of the gauge transform as described in the appendix, but for the operator $\hat H_{\tilde L}:=H+\ep V_{\tilde L}$. The main difference with the gauge transform procedure for the previous section is that now the width of each resonant zone 
decreases as $n$ increases. More precisely, we put
\bee\label{Rtheta}
R(\theta)=R_n(\theta):=\{\xi\in\R,\ |(\xi +2\theta)^2-\xi^2|<\ep_n^{1/2}\}
=(-\theta-\frac{\ep_n^{1/2}}{4|\theta|},-\theta+\frac{\ep_n^{1/2}}{4|\theta|}). 
%\ \ \ \delta=\delta(N).
\ene
and
\bee\label{phi1}
\varphi_{\theta}(\xi)=\varphi_{\theta;n}(\xi):=\varphi((\xi+\theta)4|\theta|\ep_n^{-1/2}).
%\begin{cases}
%0, & if \ \ \xi\in R(\theta) \\
%1, & \ otherwise
%\end{cases}
\ene

%with $\Theta_{\tilde L}$ instead of $\Theta$. 
Then the frequencies of the resulting operator $H_2$ will be inside the set $(\Theta_{\tilde L})_{3N}=\Theta_{3N\tilde L}$.

Note that the resonant zones obtained at each step do not intersect. 
Indeed, suppose that $\theta_1,\theta_2\in \Theta'_{3N\tilde L}$, $\theta_1\not=\theta_2$. Then $\theta_2-\theta_1\in\Theta'_{6N\tilde L}$ and, therefore, our diophantine condition implies 
\bee\label{dioph}
|\theta_2-\theta_1|\gg (6N\tilde L)^{-P_0}\gg\ep_n^{\frac{3NP_0}{P}}>\ep_n^{1/8}
\ene
for sufficiently small $\ep_n$, assuming that $P$ is chosen so large that 
\bee\label{smoothness}
\frac{3NP_0}{P}<1/8. 
\ene
At the same time the length of the resonant zone corresponding to $\theta\in\Theta'_{3N\tilde L}$ is bounded from above by $$\ep_n^{1/2}/|\theta|\le\ep_n^{1/2}(3N\tilde L)^{P_0}\ll\ep_n^{1/2}\ep_n^{-\frac{3NP_0}{P}}<\ep_n^{3/8}.$$
\ber
Of course, condition \eqref{smoothness} means that the bigger $N$ is (i.e. the more asymptotic terms we want to obtain), the bigger $P$ we should take (i.e. the smoother potentials we have to consider). 
\enr
\ber\label{wide}
The calculation above shows that even  
if we consider `wide' resonance zones which are ten times wider than \eqref{Rtheta} (i.e. `wide' resonance zones are the intervals  $(-\theta-\frac{5\ep_n^{1/2}}{2|\theta|},-\theta+\frac{5\ep_n^{1/2}}{2|\theta|})$), then these zones will not intersect either. This observation will be useful later on. 
\enr

This construction leads to two operators, $H_1$ and $H_2$ with the same properties as described in the previous section. For each $\theta\in\Theta'_{3N\tilde L}$ we denote by $R(\theta)=R(\theta;n)$ the resonant zone -- the interval centred at $-\theta$ of length $\frac{\ep_n^{1/2}}{2|\theta|}$. We also denote 
\bee
\BR(\ep_n)=\BR_n:=\cup_{\theta\in\Theta'_{3N\tilde L}}R(\theta;n);
\ene 
this is the resonant zone corresponding to $I_n$. The meaning of this set is that the symbol $h_2$ of $H_2$ is diagonal for $\xi\not\in \BR_n$. This means that all  Fourier coefficients $\hat h_2(\xi;\theta)=0$ whenever $\theta\ne 0$ and $\xi\not\in \BR_n$; our construction implies that even more is true: $\hat h_2(\xi;\theta)=0$ unless $\xi\in R(\theta;n)$. 
The main difference between the almost-periodic and quasi-periodic cases is the following: in the quasi-periodic case the resonant set was fixed for any given $N$ as $\delta(N)$ and decreasing as $N$ grows (see \eqref{BR}), whereas in the almost-periodic case $\BR(\ep_n)$ is fixed only when $\ep\in I_n$, and in general it is no longer true that $\BR_{n+1}\subset\BR_n$ (since the smaller $\ep_n$ leads to bigger $n$ and bigger $\tilde L(n)$ given by \eqref{tildeL} and, thus, $\BR_{n+1}$ consists of a bigger number of smaller zones than $\BR_n$). 
Estimate \eqref{number} implies that the number of elements in $\Theta_{3N\tilde L}$ can be estimated by
\bee
%{{3N\tilde L}\choose{3l}}
(9N\tilde L)^{3l}\ll\ep_n^{-\frac{9Nl}{P}}\ll\ep_n^{-1/6},
\ene
which implies 
\bee
\meas( \BR_n)<\ep_n^{1/6} 
\ene
if we choose $P$ large enough.

Let us now discuss the behaviour of the gaps of $H_2$ (and, therefore, of $H$). This can be done using the arguments from the quasi-periodic case.  
 When $\ep\in I_n$, the operator $H_2$ has gaps around points $|\theta|^2$, $\theta\in\Theta_{3N\tilde L(n)}$, and the length of each such gap has asymptotic expansion in natural powers of $\ep$, according to Theorem \ref{gapst}. 
Now we notice that if $\theta\in\Theta_{3N\tilde L(n)}$, then $\theta\in\Theta_{3N\tilde L(m)}$ for any $m\ge n$ and, therefore, there is a gap of $H_2$ around $\theta$ for any $m\ge n$. The length of this gap has an asymptotic expansion given by  Theorem \ref{gapst} for $\ep\in I_m$, $m\ge n$ (Here we assume that $\varepsilon_0$ is chosen to be small enough, depending only on $N$). These expansions may be different in general, but we can use Lemma \ref{main_lem} to deduce that we have a complete power asymptotic expansion of the length of a gap valid for all $\ep<\ep_0$. Thus, we obtain Theorem ~\ref{gapst1int} for smooth almost-periodic case.

%\bet \label{gapst1} Suppose, $V$ is almost-periodic and satisfies all the above assumptions.  Suppose, %$\theta\in\Theta_{\infty}'$. Then for sufficiently small $\ep$ operator $H$ has a (possibly trivial) spectral gap around %$|\theta|^2$, the length of which has  a complete asymptotic expansion in natural powers of $\ep$, as well as its %upper and lower ends. If $\hat V_\theta\not=0$ then the asymptotic expansion for the upper (lower) end of the gap %starts with $|\theta|^2\pm |\hat V_\theta|\varepsilon +O(\varepsilon^2)$.
%If $\hat h_2(-\theta_0;2\theta_0)\equiv 0$ for each $N$, then the size of the spectral gap around $|\theta|^2$ is $O(\varepsilon^{\infty})$. If $\hat h_2(-\theta_0;2\theta_0)\not\equiv 0$ for some $N$, 
%\ent

%\ber
%We cannot guarantee that such an expansion is always non-trivial, i.e. it could happen, in principle, that the length of %the gap is $O(\ep^{+\infty})$. 
%\enr

Now we discuss the asymptotic behaviour of the IDS. Recall that all our constructions are made for fixed $N$; sometimes, we will be emphasising this and make $N$ an argument of the objects we consider. First, we introduce the set of $\xi>0$ such that $\xi\not\in\Theta_\infty$ and there is an infinite sequence $n_j\to\infty$ and $\theta_j\in \Theta_{L(n_j)}$ satisfying $\xi\in R(\theta_j;n_j)$. We denote this set by $\tilde\CS_1(N)$. Since we have 
$
\sum_{n=p}^\infty\meas(\BR_n)\to 0 
$ 
as $p\to\infty$, the measure of $\tilde\CS_1(N)$ is zero. Also, it is easy to see that the set $\cap_n\BR_n(N)$ is the Cantor-type set (i.e. a perfect set with empty interior) and is, thus, uncountable (unless $V$ is periodic and $\Theta_{\infty}$ is therefore discrete). Since, obviously, $\cap_n\BR_n(N)\subset(\tilde\CS_1(N)\cup\Theta_\infty)$ and $\Theta_\infty$ is countable, this implies that the set $\tilde\CS_1(N)$ is uncountable. We also have $\tilde\CS_1(N)\subset \tilde\CS_1(\tilde N)$ for $N<\tilde N$. Finally, we introduce $\CS_1:=\cup_N\tilde\CS_1(N)$ - global uncountable set of Lebesgue measure zero. %called {\it super-resonance set}.

Let us assume at the moment that $\tau=0$.  For each fixed $\la>0$ there are the following three possibilities: 

1. Let $\sqrt{\la}\in\Theta_\infty$. Then $\sqrt{\la}=|\theta|\in R(-|\theta|;n)$ for all sufficiently large $n$ 
and we therefore can repeat the procedure from the previous Section to obtain the resonance asymptotic `expansion' \eqref{eq:Nconst1a} (see also Lemma~\ref{main_lem}).  

2. Let $\sqrt{\la}\not\in(\CS_1\cup\Theta_\infty)$. Then for all $N$ we have $\sqrt{\la}\not\in\tilde\CS_1(N)\cup\Theta_\infty$. Thus, for all sufficiently large $n$ we have $\la^{1/2}\not\in \BR_n$. Then again we can repeat the (non-resonant) procedure from the previous Section which, together with Lemma \ref{main_lem}, guarantees the existence of the complete asymptotic expansion \eqref{nonresas}. 

3. Let $\sqrt{\la}\in\CS_1$. This is the most interesting case. As we will see below, in general there is a big part of $\CS_1$ where no power asymptotic expansion exists. Let us make a pause for a moment and summarize what we have done so far. 
We have proved the following statement: 

\bet    
Suppose, $V$ is smooth %or analytic, 
almost-periodic with 
the constant Fourier coefficient $\tau=0$. Then 
there exists a set $\CS_1$  such that 
for $\la^{1/2}\in\R_{+}\setminus (\CS_1\cup \Theta_{\infty})$, we have a complete expansion of the form \eqref{nonresas}, whereas when $\la\in\Theta_{\infty}$, we have \eqref{eq:Nconst1a}. The set $\CS_1$ is uncountable and has measure zero.   
\ent

Suppose now $\tau\ne 0$. Let us denote by $R'(\theta;n)$ the interval centred at $-\theta$, but of twice larger length than $R(\theta;n)$; obviously, $R(\theta;n)\subset R'(\theta;n)$. We also denote by $\tilde\CS_2(N)$ the set of points $\xi\not\in\Theta_{\infty}$ for which there is an infinite sequence $n_j\to\infty$ and $\theta_j\in \Theta_{L(n_j)}$ such that $\xi\in R'(\theta_j;n_j)$. We put $\CS_2=\CS_2(\tau):=\cup_N\tilde\CS_2(N)$. Then $\CS_1\subset\CS_2$, $\meas(\CS_2)=0$, and for $\sqrt\la\not\in\CS_2$ we still have the complete asymptotic expansion.  Indeed, if $\ep\in I_{n_j}$ and $\sqrt\la\not\in R'(\theta_j;n_j)$, then 
$\sqrt{\la+\tau\ep}\not\in R(\theta_j;n_j)$ for sufficiently large $n$. 
This proves the following statement:

\bet
The statements of the previous theorem hold for any $\tau\not=0$ with the set $\CS_1$ replaced by a different uncountable zero measure set $\CS_2=\CS_2(\tau)$. 
\ent

Now we will prove the opposite -- that there is a substantial set $\CS$
such that for $\sqrt{\la}\in\CS$ there is no asymptotic expansion in powers of $\ep$ for $N(\la;H)$. Obviously, the measure of $\CS$ has to be zero, but we will show that it is uncountable. However, as we have seen in the previous Section, such a set must be empty in the quasi-periodic case. This means that we need to make a further assumption on the potential. Namely, we will assume that $V$ is not periodic (i.e., $\Theta_{\infty}$ is dense) and $\hat V_\theta\not=0$ for any $\theta\in\Theta'_\infty$.

\ber
We can replace the last condition by requiring that there are infinitely many non-zero Fourier coefficients located in `strategically important' places.
\enr

We again start with the case $\tau=0$. The strategy of the proof will be as follows. First, we will make a natural attempt to construct a set $\CS$ such that for $\sqrt{\la}\in\CS$ there is no asymptotic expansion in powers of $\ep$ of $N(\la;H)$. This attempt will almost work, but not quite. Then we will see what  the problem with our first attempt is and will modify it correspondingly.

So,  we define $R''(\theta;n)=(-\theta-\de_n(\theta),-\theta+\de_n(\theta))$ as the interval centred at $-\theta$ of half-length $\de_n$ and at our first attempt we define $\de_n(\theta)=\ep_n|\hat V_\theta|(100|\theta|)^{-1} $; obviously, then $R''(\theta;n)\subset R(\theta;n)$ for large $n$. Note that our constructions guarantee that if $\xi\in R''(\theta;n)$ and $\ep\in I_n$, then $|\xi|^2$ is well inside the spectral gap of $H_2(n)$ (this is the operator $H_2$, when we want to emphasize that we have performed the gauge transform for $\ep\in I_n$). 
 Now we consider the set $\tilde\CS_3(N)$ of all $\la$ for which the following two conditions are satisfied: 

a. There is an infinite sequence $n_j\to\infty$ and $\theta_j\in \Theta'_{3N\tilde L(n_j)}$ such that $\la^{1/2}\in R''(\theta_j;n_j)$, 

and

b. There is an infinite sequence $n_j'\to\infty$ such that $\la^{1/2}\not\in \BR(n_j')$. 

A simple argument based on the fact that $\Theta_{\infty}$ is dense in $\R$ implies that $\tilde\CS_3(N)$ is uncountable. 

Suppose, $\sqrt{\la}\in\tilde\CS_3(N)$. Then, if $\ep\in I_{n_j}$, the point $\la$ is in the spectral gap of $H_2(n_j)$ and, therefore, we have the (trivial) resonant asymptotic expansion \eqref{eq:Nconst}. On the other hand, if $\ep\in I_{n_j'}$, we have the non-resonant asymptotic expansion \eqref{nonresas0}. It is very tempting to stop the proof here by stating that these two expansions are different. However, we cannot quite guarantee this -- it may well happen that all the coefficients in the non-resonant  expansion \eqref{nonresas0} turn to zero. One way of overcoming this is to show that for generic set of Fourier coefficients of $V$ these coefficients are bounded away from zero. We, however, will assume a different strategy and reduce the set $\tilde\CS_3(N)$ even further (by choosing smaller values of the parameters $\de_n(\theta)$). 

Before doing this, let us see what happens with the position of the point $\xi\in \tilde\CS_3(N)$ related to different resonant zones as $n$ changes. When $n=n_j$, our point $\xi$ is inside the resonant zone  $R''(\theta_j;n_j)$ and, therefore, we have a trivial expansion for $\ep\in I_{n_j}$. If we consider values $n$ bigger than $n_j$, then $\xi$ may stay inside $R(\theta_j;n)$ for a while, but since $\cap_n R(\theta_j;n)=|\theta_j|\ne\xi$, for sufficiently large $n$ our point $\xi$ will get outside of the resonant zone $R(\theta_j;n)$; %and, moreover, outside the entire resonant set $R(n)$; 
let us denote by $\tilde k_j$ the index when this happens (i.e. $\tilde k_j$ is smallest value of $n>n_j$ for which we have $\xi\not\in  R(\theta_j;n)$). Similarly, let $k_j$ be the biggest value of $n<n_j$ for which we have $\xi\not\in  R(\theta_j;n)$. Since the width of a resonance zone shrinks by a factor  $\sqrt 2$ at each step, Remark \ref{wide} implies that $\xi$ cannot `enter' a different resonance zone immediately after `leaving' $R(\theta_j;n)$, i.e. $\xi\not\in(\BR(k_j)\cup\BR(\tilde k_j))$.  
%; since each resonant zone $R(\theta;n)$ is non-existent when 
%$Z(\theta)\geq 3N\tilde L(n;N)$ and decreases with $n$ otherwise, this in particular means that  we have $Z(\theta_j)\geq 3N\tilde L(n;N)$. 
Then by our construction we have $N$ asymptotic terms of $N(\la;H^{(\ep)})$ when $\ep\in I_{k_j}$, and the coefficient in front of $\ep^2$ is easily computable and equal to 
\begin{equation}\label{newf2}
-(2\pi\xi)^{-1}f_2(\xi,0;k_j)=(2\pi\xi)^{-1}\sum\limits_{\theta\in\Theta'_{3N\tilde L(k_j)}%,\ \theta\not\in R(\theta,k_j)
}\frac{|\hat V_\theta|^2}{|\xi+2\theta|^2-|\xi|^2}.
\end{equation}
Similarly,  we have $N$ asymptotic terms of $N(\la;H^{(\ep)})$ when $\ep\in I_{\tilde k_j}$, and the coefficient in front of $\ep^2$ equals 
\begin{equation}\label{newf2a}
-(2\pi\xi)^{-1}f_2(\xi,0;\tilde k_j)=(2\pi\xi)^{-1}\sum\limits_{\theta\in\Theta'_{3N\tilde L(\tilde k_j)}%,\ \theta\not\in R(\theta,\tilde k_j)
}\frac{|\hat V_\theta|^2}{|\xi+2\theta|^2-|\xi|^2}.
\end{equation}
Notice that the sum in \eqref{newf2a} contains more terms than \eqref{newf2}; one of the extra terms corresponds to $\theta=\theta_j$ and its modulus is at least  $\frac{\varepsilon_{\tilde k_j}^{-1/2}|\hat V_{\theta_j}|^2}{9|\theta_j|}$. The rest of the extra terms give a total contribution of $O(\varepsilon_{k_j}^{N})$. Therefore, we have 
\bee\label{newf2b}
|f_2(\xi,0;\tilde k_j)-f_2(\xi,0;k_j)|\geq\frac{\varepsilon_{\tilde k_j}^{-1/2}|\hat V_{\theta_j}|^2}{9|\theta_j|}+O(\varepsilon_{k_j}^{N}).
\ene
Now we will readjust the definition of the subset $R''$ of the resonant zone $R$ by requiring that the jump \eqref{newf2b} is at least one, which can be achieved by asking that $\varepsilon^{1/2}_{\tilde k_j}<\frac{|\hat V_{\theta_j}|^2}{18|\theta_j|}$. Another way of formulation this is requesting that if $n>n_j$ satisfies 
\bee
\varepsilon^{1/2}_{n}>\frac{|\hat V_{\theta_j}|^2}{18|\theta_j|},
\ene
then $\xi\in R(\theta_j,n)$. 
Now, we define a modified set $\tilde\CS_3(N)$ which satisfies properties a and b above, but with a modified parameter $\de_n$ defining the resonant zone $R''$ given by $\de_n(\theta)=\min\{\frac{\varepsilon_n|\hat V_\theta|}{100|\theta|},\frac{|\hat V_{\theta}|^2}{72|\theta|^2}\}$. 
The calculations just above show that if $\xi\in R''(\theta_j,n_j)$, then, assuming once again that $\varepsilon_0=\varepsilon_0(N)$ is small enough, we have: 
\bee
|f_2(\xi,0;\tilde k_j)-f_2(\xi,0;k_j)|\ge 1
\ene
and, therefore, we cannot have both these coefficients small at the same time. This shows that, indeed, we cannot have a complete power asymptotic expansion (nor even an asymptotic expansion with the remainder $o(\ep^2)$) for any $\xi\in \tilde\CS_3(N)$ with $N\ge 3$. 

 If we put 
\bee
\CS_3:=\cup_{N\geq3}\tilde\CS_3(N), 
\ene
then this is an uncountable set such that there is no complete power asymptotic expansion of $N(\lambda,H)$ for $\sqrt\la\in \CS_3$.

We have proved the following result:
\bet\label{tau0}    
Suppose, $V$ is smooth %or analytic, 
almost-periodic, but not periodic, 
the constant Fourier coefficient $\tau=0$, and $\hat V_\theta\not=0$ for any $\theta\in\Theta'_\infty$. Then 
there exists an uncountable set $\CS_3$  such that when $\la^{1/2}\in\CS_3$, there is no complete power asymptotic expansion of $N(\la;H)$; even more, if $\la^{1/2}\in\CS_3$, then no asymptotic expansion of $N(\la;H)$  %$\ep^1$ (if $\tau\ne 0$), or 
with remainder estimate $o(\ep^2)$ exists.    
\ent

Suppose now that $\tau\ne 0$.   Consider the set $\tilde\CS_3(N)$ of all $\la$ for which the following two conditions are satisfied: 

a. There is an infinite sequence $n_j\to\infty$ and $\theta_j\in \Theta'_{3N\tilde L(n_j)}$ such that $(\la+\tau\ep_{n_j})^{1/2}\in R''(\theta_j;n_j)$, 

and

b. There is an infinite sequence $n_j'\to\infty$ such that $(\la+\tau\ep_{n_j'})^{1/2}\not\in \BR(n_j')$. 

A slightly more difficult than before (but still quite elementary) argument shows that $\tilde\CS_3(N)$ is uncountable for each $\tau$. Also, similar to the case $\tau=0$, if $\ep\in I_{n_j}$, the point $\la$ is in the spectral gap of $H_2(n_j)$ and, therefore, we have the (trivial) resonant asymptotic expansion \eqref{eq:Nconst}. On the other hand, if $\ep\in I_{n_j'}$, we have the non-resonant asymptotic expansion \eqref{nonresas0}, and the coefficient at the first order term in this expression equals $-\frac{\tau}{2\pi\sqrt\la}$, which means that these two expressions are different starting with $\varepsilon$, i.e. it is enough to take $N\geq2$. Putting $\CS_3:=\cup_{N\geq2}\tilde\CS_3(N)$, we will prove the analogue of Theorem \ref{tau0} in the case $\tau\ne 0$: 
% We have proved the following result:
%\bet    
%Suppose, $V$ is smooth %or analytic, 
%almost-periodic, but not periodic, 
% and $\hat V_\theta\not=0$ for any $\theta\in\Theta_\infty$. Then 
%there exists an uncountable set $\CS_3$  such that when %$\la^{1/2}\in\CS_3$, there is no complete power asymptotic expansion of $N(\la;H)$.    
%\ent
\bet\label{taunot0}    
Suppose, $V$ is smooth %or analytic, 
almost-periodic, but not periodic, 
the constant Fourier coefficient $\tau\ne 0$, and $\hat V_\theta\not=0$ for any $\theta\in\Theta'_\infty$. Then 
there exists an uncountable set $\CS_3$  such that when $\la^{1/2}\in\CS_3$, there is no complete power asymptotic expansion of $N(\la;H)$; even more, if $\la^{1/2}\in\CS_3$, then no asymptotic expansion of $N(\la;H)$ with remainder estimate $o(\ep^1)$ exists.    
\ent

Putting all the results proved in this section together, we have proved the following: 

\bet    
Suppose, $V$ is smooth %or analytic, 
almost-periodic, but not periodic, 
 and $\hat V_\theta\not=0$ for any $\theta\in\Theta'_\infty$. Then 
there exists a set $\CS$ (which we call a super-resonance set) such that a complete power asymptotic expansion of $N(\la;H)$ exists if and only if $\la\not\in\CS$.
The set $\CS$ is uncountable and has measure zero. %The same result holds if the constant coefficient of $V$ is zero %and the set of other coefficients satisfies a certain generic condition.   
\ent
\bep
We just notice that $\CS_3\subset\CS\subset\CS_1$ for $\tau=0$ and  $\CS_3\subset\CS\subset\CS_2$ for $\tau\ne 0$. 
\enp
\ber We have called the set $\CS$ {\it the super-resonant set}. 
An interesting question which we have not studied so far is what is the dimension of this set. 
\enr

\section{Appendix}

In this appendix, we will describe the method of gauge transform. 
\subsection{Preparation}
Our strategy will be to find a unitary operator which reduces $H =
H_0+\ep \op(V)$, $H_0:=-\Delta$, to another PDO, whose symbol, essentially,
depends only on $\xi$ (notice that now we have started to distinguish between the potential $V$ and the operator of multiplication by it $\op(V)$). More precisely,
we want to find operators $H_1$ and $H_2$ with the properties discussed in Sections 3 and 4.
The unitary operator
will be constructed in the form $U = e^{i\Psi}$ with a suitable
bounded self-adjoint quasi-periodic PDO $\Psi$. This is why we
sometimes call it a `gauge transform'. It is useful to
consider $e^{i\Psi}$ as an element of the group
\begin{equation*}
U(t) = \exp\{ i \Psi t\},\ \ \forall t\in\R.
\end{equation*}
\textbf{We assume that the operator $\ad(H_0, \Psi)$ is bounded, so
that $U(t) D(H_0) = D(H_0)$}. This assumption will be justified
later on. Let us express the operator
\begin{equation*}
A_t := U(-t)H U(t)
\end{equation*}
via its (weak) derivative with respect to $t$:
\begin{equation*}
A_t = H +  \int_0^t U(-t') \ad(H; \Psi) U(t') dt'.
\end{equation*}
By induction it is easy to show that
\begin{gather}
A_1 = H + \sum_{j=1}^{\tilde k} \frac{1}{j!}
\ad^j(H; \Psi) +  R^{(1)}_{{\tilde k}+1},\label{decomp:eq}\\
R^{(1)}_{{\tilde k}+1} :=  \int_0^1 d t_1 \int_0^{t_1} d t_2\dots
\int_0^{t_{\tilde k}} U(-t_{{\tilde k}+1}) \ad^{{\tilde k}+1}(H; \Psi) U(t_{{\tilde k}+1})
dt_{{\tilde k}+1}.\notag
\end{gather}
%!!! I have found the name of this formula; the Hadamard lemma is with infinitely many terms but without the remainder
%(in fact, this formula is a version of the Hadamard lemma)
The
operator $\Psi$ is sought in the form
\begin{equation}\label{psik:eq}
\Psi = \sum_{j=1}^{\tilde k} \Psi_j,\ \Psi_j = \op(\psi_j),
\end{equation}
with some bounded operators $\Psi_j$. Substitute this formula in
\eqref{decomp:eq} and rewrite, regrouping the terms:
\begin{gather}
A_1 = H_0 + \ep\op(V) + \sum_{j=1}^{\tilde k} \frac{1}{j!} \sum_{l=j}^{\tilde k} \sum_{k_1+
k_2+\dots + k_j = l}
\ad(H; \Psi_{k_1}, \Psi_{k_2}, \dots, \Psi_{k_j})\notag\\
 +
 R^{(1)}_{{\tilde k}+1} + R^{(2)}_{{\tilde k}+1},\notag\\
R^{(2)}_{{\tilde k}+1}: = \sum_{j=1}^{\tilde k} \frac{1}{j!}\sum_{k_1+ k_2+\dots + k_j
\ge  {\tilde k}+1} \ad(H; \Psi_{k_1}, \Psi_{k_2}, \dots, \Psi_{k_j}).
\label{rtilde:eq}
\end{gather}
Changing this expression yet again produces
\begin{gather*}
A_1 = H_0 + \ep\op(V) + \sum_{l=1}^{\tilde k} \ad(H_0; \Psi_l) + \sum_{j=2}^{\tilde k}
\frac{1}{j!} \sum_{l=j}^{\tilde k} \sum_{k_1+ k_2+\dots + k_j = l}
\ad(H_0; \Psi_{k_1}, \Psi_{k_2}, \dots, \Psi_{k_j})\\
+ \sum_{j=1}^{\tilde k} \frac{1}{j!} \sum_{l=j}^{\tilde k} \sum_{k_1+ k_2+\dots + k_j
= l} \ad(\ep\op(V); \Psi_{k_1}, \Psi_{k_2}, \dots, \Psi_{k_j}) +
R^{(1)}_{{\tilde k}+1} + R^{(2)}_{{\tilde k}+1}.
\end{gather*}
Next, we switch the summation signs and decrease $l$ by one in the
second summation:
\begin{gather*}
A_1 = H_0 + \ep\op(V) + \sum_{l=1}^{\tilde k} \ad(H_0; \Psi_l) + \sum_{l=2}^{\tilde k}
\sum_{j=2}^l \frac{1}{j!}\sum_{k_1+ k_2+\dots + k_j = l}
\ad(H_0; \Psi_{k_1}, \Psi_{k_2}, \dots, \Psi_{k_j})\\
+\sum_{l=2}^{{\tilde k}+1} \sum_{j=1}^{l-1} \frac{1}{j!}\sum_{k_1+ k_2+\dots
+ k_j = l-1} \ad(\ep\op(V); \Psi_{k_1}, \Psi_{k_2}, \dots, \Psi_{k_j}) +
R^{(1)}_{{\tilde k}+1} + R^{(2)}_{{\tilde k}+1}.
\end{gather*}
Now we introduce the notation
\begin{gather}
B_1 := \ep\op(V),\notag\\
B_l := \sum_{j=1}^{l-1} \frac{1}{j!}
 \sum_{k_1+ k_2+\dots + k_j = l-1}
\ad(\ep\op(V); \Psi_{k_1}, \Psi_{k_2}, \dots, \Psi_{k_j}),
\ l\ge 2,\label{bl:eq}\\
T_l := \sum_{j=2}^l \frac{1}{j!} \sum_{k_1+ k_2+\dots + k_j = l}
\ad(H_0; \Psi_{k_1}, \Psi_{k_2}, \dots, \Psi_{k_j}),\  l\ge 2.
\label{tl:eq}
\end{gather}
We emphasise that the operators $B_l$ and $T_l$ depend only on
$\Psi_1, \Psi_2, \dots, \Psi_{l-1}$. Let us make one more
rearrangement:
\begin{gather}
A_1 =  H_0 + \ep\op(V) + \sum_{l=1}^{\tilde k} \ad(H_0, \Psi_l) + \sum_{l=2}^{\tilde k} B_l
+   \sum_{l=2}^{{\tilde k}} T_{l} + R_{{\tilde k}+1},\notag\\
R_{{\tilde k}+1} =  B_{{\tilde k}+1} + R^{(1)}_{{\tilde k}+1} + R^{(2)}_{{\tilde k}+1}.\label{r:eq}
\end{gather}

Let $\varphi_{\theta}(\xi,\ep_n)$ be a smooth cut-off function of the set 
\begin{equation}\label{setdef}\xi:\ ||\xi+2\theta|^2-|\xi|^2|>\ep_n^{1/2},\ \ \ \varepsilon_n:=2^{-n}\varepsilon_0%\ \hbox{and}\ |\xi+\theta|^2<2\lambda
.\end{equation}
More precisely, let $\psi=\psi(\xi)$ be a standard smooth non-negative cut-off function satisfying 
$\supp \psi \subset [-1/2,1/2]$ and $\psi(\xi)=1$ for $\xi\in[-1/4,1/4]$, and let   
$\varphi:=1-\psi$. 
We put 
\bee\label{phi1a}
\varphi_{\theta}(\xi,\ep_n):=\varphi((\xi+\theta)4|\theta|\ep_n^{-1/2}),\ \ \ \varepsilon_n:=2^{-n}\varepsilon_0.
\ene
For any symbol $$b:=\sum_{\theta}\hat b(\xi,\theta)\be_{2\theta}(x)$$ we use the notation
$$b^{\natural}=b^{\natural}(\ep_n):=\sum_{\theta} \hat b(\xi,\theta)\varphi_{\theta}(\xi,\ep_n)\be_{2\theta}(x).$$ Similar notation is used for corresponding operator, i.e. $B^{\natural}$.

Now we can specify our algorithm for finding $\Psi_j$'s. The symbols
$\psi_j$ will be found from the following system of commutator
equations:
\begin{gather}
\ad(H_0; \Psi_1) + B_1^{\natural} = 0,\label{psi1:eq}\\
\ad(H_0; \Psi_l) + B_l^{\natural} + T_l^{\natural} = 0,\ l\ge
2,\label{psil:eq}
\end{gather}
and hence
\begin{equation}\label{lm:eq}
\begin{cases}
A_1 = H_0 + Y_{\tilde k}-Y^\natural_{\tilde k} + R_{{\tilde k}+1},\\[0.3cm]
Y_{\tilde k} =  \sum_{l=1}^{{\tilde k}} B_l + \sum_{l=2}^{{\tilde k}}
T_l.
\end{cases}
\end{equation}
Below we denote by $y_{\tilde k}$ the symbol of the PDO $Y_{\tilde k}$. Obviously, the operators $B_l^{\natural},
T_l^{\natural}$ are bounded, and therefore, in view of
\eqref{psi1:eq}, \eqref{psil:eq}, so is  the commutator $\ad(H_0;
\Psi)$. This justifies the assumption made in the beginning of the
formal calculations in this section.

It is also convenient to introduce the following norm in the class of symbols:
\begin{equation}\label{symbolnorm}
\1 b\1:=\sum_{\theta}\sup_{\xi}|\hat b(\xi,\theta)|.
\end{equation}
We notice that $\|\op(b)\|\leq \1 b\1$.

\subsection{Commutator equations}

Put
$$
\tilde{\chi}_{\theta}(\xi):=\varphi_{\theta}(\xi)(|\xi+2\theta|^2-|\xi|^2)^{-1}=\frac{\varphi_{\theta}(\xi)}
{4(\xi+\theta)\theta}
$$
when $\theta\not={0}$, and $\tilde{\chi}_{0}(\xi)=0$.
We have
\begin{lem} \label{commut:lem}
Let $A = \op(a)$ be a symmetric PDO with symbol $a$ such that $\1 a\1<\infty$. Then the
PDO $\Psi$ with the Fourier coefficients of the symbol $\psi(
\xi,x)$ given by
\begin{equation}\label{psihat:eq}
\hat\psi(\xi,\theta) = i\,{\hat a}(\xi,\theta)\tilde{\chi}_{\theta}(\xi)
\end{equation}
solves the equation
\begin{equation}\label{adb:eq}
\ad(H_0; \Psi) + \op(a^{\natural})= 0.
\end{equation}
Moreover, the operator $\Psi$ is bounded and self-adjoint and its symbol $\psi$ satisfies the
following bound:
\begin{equation}\label{psitau:eq}
\1\psi\1\leq \ep_n^{-1/2}\1 a\1.
\end{equation}
%where
%\begin{equation}\label{sigma:eq}
%\s = \om - (2m - 2)\b^{-1}-1.
%\end{equation}
\end{lem}
Now, using  Lemma~\ref{commut:lem}, equations \eqref{setdef}, \eqref{psi1:eq}, \eqref{psil:eq} and applying inductive arguments (cf. the proof of Lemma 4.2 from \cite{ParSob}), we obtain the following estimates for the symbols introduced above:
\bel\label{estimateskm:lem}
Let $V$ be a symmetric symbol. Then $\psi_j,\, b_j,\,t_j$ satisfy (for $\varepsilon\leq\varepsilon_n$)
\bee\label{estpsi}
\1\psi_j\1\leq C_j\ep_n^{\frac12 j}\left(\1 V\1\right)^j,\ \ j\geq 1;
\ene
\bee\label{estbt}
\1 b_j\1+\1 t_j\1\leq C_j\ep_n^{\frac12 j +\frac12}\left(\1 V\1\right)^j,\ \ j\geq 2.
\ene
Moreover, assuming $\ep_0$ is small enough (depending on $V$ and ${\tilde k}$) we get
\bee\label{estpsitotal}
\1\psi\1\ll\ep_n^{1/2}\1 V\1;
\ene
\bee\label{esty}
\1 y_{\tilde k}\1\leq 2\ep_n\1 V\1;
\ene
\bee\label{esterror}
\| R_{{\tilde k}+1}\|\ll \ep_n^{\frac12 \tilde k +1}\left(\1 V\1\right)^{{\tilde k}+1}.
\ene
\enl
\bep The proof follows by induction. For $B_1$ the estimate follows from the definition and for $\Psi_1$ from Lemma~\ref{commut:lem}. Now, fix $l\geq1$ and assume \eqref{estpsi} and \eqref{estbt} for all $l\geq j\geq 1$. From \eqref{bl:eq} we get
\begin{equation}\label{raz}
\1 b_{l+1}\1\leq C_{l+1} \ep_n^{1+\frac12 l }\left(\1 V\1\right)^{l+1}= C_{l+1}\ep_n^{\frac12 (l+1) +\frac12}\left(\1 V\1\right)^{l+1}.
\end{equation}
Next, we use definition \eqref{tl:eq}. For the first commutator we apply \eqref{psi1:eq} or \eqref{psil:eq}. Then assumption of the induction implies
\begin{equation}\label{dva}
\1 t_{l+1}\1\leq C_{l+1} \ep_n^{\frac12 k_1 +\frac12 +\frac12 (l+1-k_1)}\left(\1 V\1\right)^{l+1}= C_{l+1}\ep_n^{\frac12 (l+1) +\frac12}\left(\1 V\1\right)^{l+1}.
\end{equation}
This proves \eqref{estbt} for $j=l+1$. Aplying Lemma~\ref{commut:lem} we get \eqref{estpsi} for $j=l+1$. 
\enp

Now, we take
\bee\label{eq:kM}
{\tilde k}>2N.
\ene
%and assume that $k$ is large enough so that $\ep_n^{(\tilde k+1)^20-}\ll\ep_n^{-3/4}$. 
Then
$$
\| R_{{\tilde k}+1}\|\ll\ep_n^{N}
$$
and we can disregard $R_{{\tilde k}+1}$. More precisely, let $W=W_{\tilde k}$ be the operator with symbol
\bee\label{eq:newy}
w_{\tilde k}(\xi,x):=y_{\tilde k}(\xi,x)-y_{\tilde k}^{\natural}(\xi,x),\ \ \hbox{i.e.}\ \
\hat w_{\tilde k}(\xi,\theta)=\hat y_{\tilde k}(\xi,\theta)(1-\varphi_{\theta}(\xi)).
\ene
We put $H_1:=A_1$ and $H_2:=-\Delta+W$. Then $||H_1-H_2||\ll\ep_n^{N}$ and, moreover, the symbol $\hat h_2(\xi,\theta):=\xi^2+\hat w(\xi,\theta)$ satisfies conditions described in Sections 3 and 4. 

\ber\label{quasiperiodic}
In the case of quasi-periodic potential $V$ the construction above can be simplified. In fact, we can use just $\delta(N)$ instead of $\varepsilon_n^{1/2}$ (see Section 3) and thus, avoid further "glueing" of the asymptotics for different intervals $I_n$ (see Lemma~\ref{main_lem}). Technically, it is possible because different resonant zones do not intersect for all steps up to $9N$ and thus, the cut-off functions $\varphi_\theta(\xi)$ present in the symbol of the operator $H_2$ (see Lemma~\ref{lem:symbol} below) are equal either to $0$ or $1$ for all $\xi$ close to the center of a given resonant zone or between any two neighboring resonant zones. This ensures analyticity of the functions used in the proof of $\eqref{nonresas0}$ (for nonresonant $\xi$) and Lemma~\ref{gaps} and asymptotics below (for resonant $\xi$).
\enr

\subsection{Computing the symbol of the operator after gauge transform}
The following lemma provides us with more explicit form of the
symbol ${y}_{\tilde k}$. 
\bel\label{lem:symbol} We have $\hat{y}_{\tilde k}(\xi,\theta)=0$ for $\theta\not\in\Theta_{\tilde k}$. Otherwise,
\begin{equation}\label{symbol}
\begin{split}
&\hat{y}_{\tilde k}(\xi,\theta)=\ep\hat{V}_\theta+\sum\limits_{s=1}^{{\tilde k}-1}\ep^{s+1}\sum  C_s(\xi,\theta)\hat{V}_{\theta_{s+1}}\prod\limits_{j=1}^s \hat{V}_{\theta_j}\tilde{\chi}_{\theta_j'}(\xi+2\phi_j')\cr
&=\ep\hat{V}_\theta+\sum\limits_{s=1}^{{\tilde k}-1}\ep^{s+1}\sum
C_s(\xi,\theta)\hat{V}_{\theta_{s+1}}\prod\limits_{j=1}^s \hat{V}_{\theta_j}\frac{\varphi_{\theta_j'}(\xi+2\phi_j')}
{4\theta_j'(\xi+2\phi_j'+\theta_j')},
\end{split}
\end{equation}
where the second sums are taken over all $\theta_j\in\Theta$,
$\theta_j',\phi_j'\in\Theta_{s+1}$ and
\begin{equation}\label{indsconst}
C_s(\xi,\theta)=\sum\limits_{p=1}^s\sum\limits_{\theta_j '',\phi_j
''\in\Theta_{s+1}\ (1\leq j\leq p)}C_s^{(p)}(\theta) \prod\limits_{j=1}^p
\varphi_{\theta_j ''}(\xi+2\phi_j '').
\end{equation}
Here $C_s^{(p)}(\theta)$ depend on $s,\ p$ and all vectors
$\theta,\theta_j,\theta_j',\phi_j',\theta_j '',\phi_j ''$. At the
same time, coefficients $C_s^{(p)}(\theta)$ can be bounded uniformly
by a constant which depends on $s$ only. We apply the convention that $0/0=0$.
\enl
The proof is identical to the proof of Lemma 9.3 from \cite{PaSh1} and we omit it here. Explicit value of the coefficients for the second term (see \eqref{eq:nonres2} and \eqref{eq:nonres2a}) can be found directly as the second order perturbation or following more carefully the first two steps of the construction for $A_1$ from \eqref{lm:eq}.

\bibliographystyle{amsplain}

\begin{thebibliography}{10}

\bibitem{A} V.I.Arnol'd, \emph{Remarks on the perturbation theory for problems of Mathieu type}, Russian Mathematical Surveys, \textbf{38}(4) (1983), 215--233. 

\bibitem{Bor} D.I.Borisov, \emph{On the spectrum of the Schr\"odinger
operator perturbed by a rapidly oscillating potential}, Journal of Mathematical Sciences, \textbf{139}(1), 2006, 6243-6322.  


\bibitem{DaGo} D. Damanik, M. Goldstein, \emph{ On the inverse spectral problem for the quasi-periodic Schr\"odinger
equation}, Publ. Math. Inst. Hautes Etudes Sci., \textbf{119}(1) (2014), 217--401.

\bibitem{Le} M. Leguil, J. You, Z. Zhao, Q. Zhou, \emph {Asymptotics of spectral gaps of quasi-periodic Schrödinger operators},
 preprint arXiv:1712.04700, 2017.

\bibitem{LiSh} W. Liu, Y. Shi, \emph{Upper bounds on the spectral gaps of quasi-
periodic Schr\"odinger operators with Liouville frequencies}, preprint arXiv:1708.01760, 2017.


\bibitem{MPSh} S. Morozov, L. Parnovski, R. Shterenberg, \emph{Complete Asymptotic Expansion of the Integrated Density of States of Multidimensional Almost-Periodic Pseudo-Differential Operators}, Annales Henri Poincar\'e, \textbf{15}(2) (2014), 263--312. 

\bibitem{PaSh1} L. Parnovski, R. Shterenberg, \emph{Complete asymptotic expansion of the integrated density of states of multidimensional almost-periodic Schr\"odinger operators}, Ann. of Math.,  \textbf{176} (2) (2012), 1039--1096.

\bibitem{ParSob} L. Parnovski, A. Sobolev, \emph{Bethe-Sommerfeld conjecture for periodic operators with strong perturbations}, Inv. Math., \textbf{181}(3) (2010), 467--540.

\end{thebibliography}

\providecommand{\bysame} {\leavevmode\hbox
to3em{\hrulefill}\thinspace}

\end{document}